\documentclass[pra,aps]{revtex4}
\pdfoutput=1
\usepackage{amssymb}
\usepackage{amsfonts} %paquete para escribir letras como el cuerpo de los reales
\usepackage{amsthm} %paquete matemÃ¡tico
\usepackage{amsmath} %paquete matemÃ¡tico
\usepackage{hyperref}
\usepackage{bbold}
 \usepackage{xcolor}
 \usepackage{graphicx}
\usepackage{rotating}
\newtheorem{theorem}{Theorem}

\newtheorem{corollary}[theorem]{Corollary}

\newtheorem{proposition}[theorem]{Proposition}

\newcommand{\be}{\begin{equation}}
\newcommand{\ee}{\end{equation}}
\newcommand{\bea}{\begin{eqnarray}}
\newcommand{\eea}{\end{eqnarray}}
\newcommand{\bd}{\begin{displaymath}}
\newcommand{\ed}{\end{displaymath}}

\newcommand{\ad }{{\hat{a}}^{\dagger}}

\newcommand{\ah }{ \hat{a}}

\def \um {\frac{1}{2}}

\begin{document}

%\draft
%\preprint{
%\begin{tabular}{l}
%\hbox to\hsize{\hfill KAIST-TH 2006/10}\\
%[-1mm]
%\hbox to\hsize{\hfill KIAS-P06047}\\
%[-2mm] \hbox to\hsize{\hfill hep-ph/yymmdd}\\
%[-3mm] \hbox to\hsize{\hfill November 2011}\\
%[-3mm]
%\end{tabular}
%}

\title{Quantum statistical properties of multiphoton hypergeometric coherent states and the discrete circle representation}
\author{S. Arjika}
\affiliation{Department of Mathematics and Computers Sciences, University of Agadez, Niger}
\email{rjksama@univ-agadez.edu.ne}
\author{M. Calixto}
\affiliation{Department of Applied Mathematics and Institute Carlos I of Theoretical and Computational Physics, University of  Granada,
Fuentenueva s/n, 18071 Granada, Spain}
\email{calixto@ugr.es}
\author{J. Guerrero}
\affiliation{Department of Mathematics, University of Jaen, Campus Las Lagunillas s/n, 23071 Jaen, Spain}
\email{jguerrer@ujaen.es}

\date{\today}

\begin{abstract}
We review the definition of hypergeometric coherent states, discussing some representative examples. Then we study mathematical and statistical properties of hypergeometric Schr\"odinger cat states, defined as orthonormalized eigenstates of 
$k$-th powers of nonlinear $f$-oscillator annihilation operators, with $f$ of hypergeometric type. 
These ``$k$-hypercats''  can be written as an equally weighted superposition of hypergeometric coherent states $|z_l\rangle, l=0,1,\dots,k-1$, with $z_l=z e^{2\pi i l/k}$ a $k$-th root of  $z^k$, and they interpolate between number and coherent states. 
This fact motivates a continuous circle representation for 
high $k$. We also extend our study to truncated hypergeometric functions (finite dimensional Hilbert spaces) and a discrete exact circle representation is provided. We also show how to generate  $k$-hypercats by 
amplitude dispersion in a Kerr medium and analyze their generalized Husimi $Q$-function in the super- and sub-Poissonian cases at different fractions of the revival time.
\end{abstract}

%\pacs{PACS numbers:02.20.Uw; 05.30.-d; 05.30.Jp }

\maketitle

\noindent\textbf{Keywords:} nonlinear, nonclassical, macroscopic superpositions of quantum states, generalized even/odd states, multicomponent Schr\"odinger cat states, multiphoton or circular states.\\
\noindent\textbf{MSC:} 81R30

\section{Introduction}
The subject of Coherent States (CS) is traced back to 1926 when  Schr\"odinger first introduced the notion of (canonical) CS of the harmonic oscillator \cite{Schrodinger}. Later, R. Glauber  \cite{R.Glauber} realized the importance of CS 
in the description of the radiation field. Since then, the subject of CS has grown and pervades almost all branches of quantum physics (see e.g. \cite{Klauder,W.Zhang&D.Feng&R.Gilmore} and \cite{JPAissue,Vourdas} for old and recent reviews and  
\cite{KlauderS,Perelomovbook,Dodonovbook,Gazeaubook,Glauber} for standard textbooks). Besides, some other important topics in applied mathematics, like the theory of wavelets, 
are also related to the notion of CS \cite{S.Ali&J.Antoine&J.Gazeau}. Standard (canonical) CS have been generalized in many ways. For example, in 1972, Gilmore \cite{Gilmore1,Gilmore11} and Perelomov \cite{Perelomov,Perelomovbook} 
realized that canonical CS were rooted in group theory (the Heisenberg-Weyl group)  and generalized the notion of CS by extending the concept of displacement operator $D(z)=\exp(z\ah-\bar z\ad)$ for other types of Lie groups. 
Typical examples are spin-$s$, Bloch, $SU(2)$ or atomic CS, whose properties where studied by Radcliffe \cite{Radcliffe71}, Gilmore \cite{Gilmore1,Gilmore11} and Perelomov \cite{Perelomov}. 
Also, the concept of squeezing is closely linked to the non compact $SU(1,1)$ group. 
In general, there are several definitions or approaches to CS, namely:
\begin{enumerate}
 \item Barut-Girardello \cite{BarutGirardello}: eigenstates of the annihilation operator.
 \item Gilmore-Perelomov: group-theoretical approach \cite{Perelomovbook}.
 \item Minimal uncertainty and intelligent states \cite{Aragone,DeMeyer}.
 \item Gazeau-Klauder \cite{GazeauKlauder}: ``non-spreading, temporally stable'', intimately  related to a Hamiltonian model.
\end{enumerate}
All definitions are equivalent for canonical CS, but this is not true in general. In this article we shall adopt the Barut-Girardello approach. In particular, we shall deal with special types of nonlinear CS \cite{fosci} related to 
the so-called $f$-oscillator annihilation and creation operators

\begin{eqnarray}
\ah_f&=&\ah f(\hat n)=f(\hat n+1)\ah=\sum_{n=0}^\infty \sqrt{n+1}f(n+1)|n\rangle\langle n+1|\nonumber\\   
\ad_f&=&f^\dag(\hat n)\ad=\ad f^\dag(\hat n+1)=\sum_{n=0}^\infty \sqrt{n+1}\bar f(n+1)|n+1\rangle\langle n| \,, \label{af}
\end{eqnarray}
where $f$ is an arbitrary function of the number operator $\hat n=\ad\ah$. 
Therefore,  the nonlinear Hamiltonian $H_f={\omega}\ad_f\ah_f$ (we use $\hbar=1$ throughout the article)  has eigenvalues 
$E_n={\omega}n |f(n)|^2,\,n=0,1,2,\ldots$, (non-equidistant energy levels, in general). This can be interpreted as an amplitude dependent frequency. 
Nonlinear $f$-CS $|z,f\rangle$ are then defined as eigenstates of $\ah_f$, i.e., $\ah_f|z,f\rangle=z|z,f\rangle$, which lead to 
\begin{equation}
 |z;f\rangle=\mathcal{N}_f^{-\um}(|z|)\sum_{n=0}^\infty \frac{z^n}{\sqrt{n!} f(n)!}|n\rangle\,,\label{nonlinearfCS}
\end{equation}
 with $f(n)!=f(1)\dots f(n)$, $f(0)!=1$ and $\mathcal{N}_f(|z|)=\sum_{n=0}^\infty \frac{|z|^{2n}}{{n!} |f(n)!|^2}$ a normalization factor. Nonlinear $f$-CS are not orthogonal in general, but they 
 form an overcomplete set and close a resolution of the identity [see later on eq. \eqref{resolhCS} for the hypergeometric case]. 
 
On the mathematical side, the general case was studied by Klauder and Penson in \cite{Klauder&PensonS}, where they constructed $f$-CS and closure relations through solutions of Stieltjes and Hausdorff moment problems  
(see \cite{Akhiezer,Tamarkin,Simon} for standard references on the moment problem). They also studied the reverse way, that is,  how to define $f$-CS
given a Hamiltonian $H$ with (non necessarily equidistant) spectrum $E_n$. On the physical side, the first proposal to generate $f$-CS was given in \cite{MatosVogel} as emergent stationary states of the motion of an
appropriately laser-driven trapped ion. Later, many other generation schemes of $f$-CS have been explored, for example: single-atom lasers, micromaser under the intensity-dependent Jaynes-Cummings model, 
excitons in a wide quantum dot, or using a mechanical resonator in an optomechanical microcavity (see e.g \cite{YanZhuLi} and references therein).

In this article we shall restrict ourselves to the special case  when $|f(n)|^2$ is a rational function of $n$, like in Eq. \eqref{fhypergeom}. 
For positive parameters $\alpha_j$ and $\beta_j$, these are related to the so called hypergeometric CS \cite{Appl,Dehghani,PopovPS} (HCS for short), 
of which we make a brief introduction and discuss some representative examples in Section \ref{hypersec}. We shall not restrict ourselves 
to positive $\alpha_j$ and $\beta_j$, but we shall also consider negative integer cases, for which $|f(n)|^2$ has 
either zeros or poles. This case leads to ``truncated HCS'', of which we also give some interesting examples in Section 
\ref{hypersec2}.

CS are said ``quasi-classical'' and they are used in Quantum Mechanics and Quantum Field Theory to study the classical limit. 
CS accurately describe the physical properties of many macroscopic quantum systems like in: quantum optics, 
Bose-Einstein condensates (BEC), superconductors, superfluids, quantum Hall effects, etc. In particular, the ground state of many physical 
systems undergoing a quantum phase transition is well described by a CS. Actually, it was Gilmore who introduced an algorithm \cite{Gilmore2}, which makes use of CS as variational
states to approximate the ground state energy, to study the classical, thermodynamic or mean-field limit of some algebraic quantum models. This algorithm has proved to be specially
suitable to analyze the phase diagram of Hamiltonian models undergoing a quantum phase transition like: the Dicke model 
of atom-field interactions  \cite{Dicke,CastaPRA,DickeEPL,DickePRA}, Bose-Einstein condensates \cite{AnalCasta},  the Lipkin-Meshkov-Glick model \cite{LMG,CastaPRB,LipkinPS,LipkinEPL}, vibron model for molecules 
\cite{vibroncurro,VibronHusimiPRA,EntangVibronPRA,parityvibron}, bilayer quantum Hall systems \cite{bilayerPRB,husiBLQH}, etc. In some quantum phases, the ground state is in fact a (parity) symmetry adapted CS or 
``Schr\"odinger cat'', $|z,{\pm}\rangle\propto|z\rangle\pm |-z\rangle$ (even and odd),  in the sense of a quantum superposition of two semi-classical (macroscopic) CS  with negligible overlap $|\langle z|-z\rangle|\ll 1$ (for large $|z|$), 
exhibiting squeezing and delocalization, among many other interesting properties. 
The idea of the even and odd CS was first introduced 
by Dodonov, Malkin and Man'ko  \cite{Dodonovcat,Mankocat} and later extended to more general finite groups than the parity group $\mathbb{Z}_2=\{1,-1\}$ \cite{Mankocat}. Nieto and Traux 
\cite{M.Nieto&D.Traux} showed that these  states  are a special set of nonclassical states and their statistical properties 
 were studied by  \cite{V.Buzek&A.Viiella-Barranco&P.Knight,M.Hillery}, among many others. Parity-adapted CS  were generalized to the nonlinear case of even and odd $f$-CS \cite{Mancini,Sivakumar}, 
 their non-classical properties depending on the induced nonlinearity $ f(\hat n)$.

Parity-adapted (even and odd) CS $|z,\pm\rangle$ are also eigenstates of $\ah^2$, so that $\ah^2|z,\pm\rangle=z^2|z,\pm\rangle$. The operator $\ah^2$ plays a fundamental role in 
creating squeezing, through the squeeze operator $S(\zeta)=\exp[\zeta \ah^2-\bar{\zeta} \hat{a}^{\dag 2}]$, and in describing the second harmonic generation (or ``frequency doubling'') in nonlinear optics and laser industry. 
The extension to $k$-th order harmonic generation and their higher-order squeezing was studied in \cite{PRA44,PRA46} . 
Eigenstates of cubic, and higher powers $\ah^k$, of the annihilation  operator lead 
to generalizations of parity-adapted (even and odd $\mathbb{Z}_2$) Schr\"odinger cat states, sometimes denoted by Schr\"odinger kittens \cite{Dodonovbook}.  However, 
the $k$-th order (multiphoton) squeeze operator  $S(\zeta)=\exp[\zeta \ah^k-\bar{\zeta}  \hat{a}^{\dag k}]$ is ill-defined for $k\geq 3$ \cite{PRD29}. 
These eigenstates of  $\ah^k$ can also be written as quantum mechanical superpositions of macroscopically 
distinguishable CS (finite superpositions of CS first appeared in \cite{Titulaer,Birula,Stoler}). They can be generated via amplitude dispersion \cite{PRL57,Kirchmair} and 
are used in quantum information processing \cite{nature06} and quantum spectroscopy \cite{nature11}. Actually, 
there is a close relation between eigenstates of  $\ah^k$ and the circle representation for CS \cite{PRA48,PRA50,PRA53,JPA31}, not to be confused with the subject of CS on the circle, of which 
we comment in appendix \ref{CScircle}. In particular, in \cite{PRA50} it was proved that standard number eigenstates $|n\rangle$ can be 
represented as a continuous superposition of CS on the circle.  More precisely, denoting $z=re^{i\theta}$, one has
\be
|n\rangle=\frac{e^{r^2/2}}{2\pi}\sqrt{n!}\, r^{-n}\int_0^{2\pi} e^{-in\theta}|r e^{i\theta}\rangle d\theta\,. 
\label{circlerepcan}
\ee
This equality is just a consequence of the analytical nature of CS and can be extended to all nonlinear CS (Cauchy theorem).  
Eigenstates of  $\ah^k$ can be seen as a discretization of the previous integral and therefore as an approximation to number states $|n\rangle$ by CS superpositions on the circle. This fact 
was exploited in \cite{samplingsphere,samplingH,samplingHW} to formulate sampling theorems and discrete Fourier transforms on the sphere, hyperboloid and complex plane, by using the circle representation 
of $SU(2)$, $SU(1,1)$ and canonical CS. In this article we want to extend all these interesting constructions to general hypergeometric-like CS. Orthonormalized eigenstates of $\ah_f^k$ where introduced by \cite{LiuJPA32},  
for general $f$, and some of its statistical properties where discussed in \cite{JPB35}. Here we shall explore many of their interesting properties for $f$ given in \eqref{fhypergeom}.

The organization of the paper is as follows. Firstly, in Section \ref{hypersec} we briefly remind the definition of HCS and their properties, the duality property and the truncation operation when 
$f(n)$ has either zeros or poles, providing numerous interesting examples. In Section \ref{kcatsec} we introduce the notion of $k$-hypercats as orthonormalized eigenstates of $\hat{a}_f^k$ and we discuss some of its 
statistical properties, which reveal that $k$-hypercats interpolate between number and coherent states. The structure of $k$-hypercats, as an equally-weighted superposition of HCS uniformly distributed 
on the circle, suggests a circle representation of number states in terms of HCS on the circle, of which $k$-hypercats constitute a finite discrete approximation. The case of truncated 
$k$-hypercats requires a different definition and they are introduced in Section \ref{hyperlikecats}, for which an exact discrete circle representation is possible. In Section \ref{kerrsec} we 
generate equally weighted, uniformly distributed on a circle, multicomponent HCS by temporal evolution in a Kerr medium. We corroborate the multicomponent structure by representing the generalized Husimi $Q$-function 
[see eq. \eqref{husifunc} for a formal definition]  in phase space and give a rough estimate for the number of distinguishable components as a function of the standard  deviation $\sigma$ and the initial displacement $|z|$. Finally, Section \ref{conclusec} is left for 
conclusions and appendices \ref{CScircle}, \ref{apA} and \ref{meannf} for some clarifications and  cumbersome formulas. We also provide a summary table (last page) with numerous interesting examples of hypergeometric CS and their duals.

\section{\label{hypersec}Hypergeometric-like CS and some representative examples}

In this article we shall consider the case
 \be
f(n)=\left(\frac{ (\beta_1+n-1) \cdots (\beta_q+n-1) }{(\alpha_1+n-1) \cdots (\alpha_p+n-1) }\right)^{\um}.\label{fhypergeom}
\ee
For positive $\alpha_j$ and $\beta_j$, the nonlinear $f$-CS of equation \eqref{nonlinearfCS} 
are related to the so called HCS \cite{Appl,Dehghani,PopovPS}, of which we give a brief in Section \ref{hypersec1} to set notation and remind their main properties.  
For negative $\alpha_j$ and/or $\beta_j$, the function $f(n)$ has zeros and poles and summations have to be truncated, which implies to deal with 
finite-dimensional Hilbert spaces (see Section \ref{hypersec2} for more details).

\subsection{\label{hypersec1}Hypergeometric CS} 

For $f$ given in  \eqref{fhypergeom}, we prefer to write nonlinear $f$-CS \eqref{nonlinearfCS} as 
\be |z;\alpha,\beta\rangle={}_pF_q(\alpha,\beta;|z|^2)^{-\um}\sum_{n=0}^\infty 
\frac{z^n}{\sqrt{{}_p\rho_q(n)}}|n\rangle,\quad {}_p\rho_q(n)\equiv  n! f(n)!^2.\label{hypercs}\ee 
where the normalization function $\mathcal{N}_f(|z|)$ is nothing but the generalized hypergeometric function 
\be
\mathcal{N}_f(|z|)={}_pF_q(\alpha,\beta;|z|^2)=\sum_{n=0}^\infty \frac{(\alpha_1)_n \cdots (\alpha_p )_n }{ (\beta_1)_n \cdots (\beta_q)_n }\frac{|z|^{2n}}{n!},\label{normfact}
\ee
where $\alpha=(\alpha_1,\dots,\alpha_p)$ and $\beta=(\beta_1,\dots,\beta_q)$ and $(\alpha)_n=\alpha(\alpha+1) \cdots (\alpha + n-1) , ~~~ (\alpha)_0 =1$ 
is the Pochhammer-symbol. The series converges for any finite $|z|$ if $p<q+1$, whereas $p=q+1$ requires in general $|z|<1$. The last condition can be relaxed to  $|z|=1$ when 
$\eta=\Re(\sum_{j=1}^p \alpha_j-\sum_{j=1}^q \beta_j)<0$ or $0\leq\eta<1$ if $z\not=1$ \cite{Appl}. 

Note that hypergeometric $f$-CS \eqref{hypercs} can also be formally written as
\be
|z;\alpha,\beta\rangle={}_pF_q(\alpha,\beta;|z|^2)^{-\um}{}_pF_q(\alpha,\beta;z \ad)|0\rangle,\label{expoF}
\ee
which resembles the usual formula $|z\rangle=\exp(-|z|^2/2)\exp(z\ad)|0\rangle$ for canonical CS, replacing the exponential by the hypergeometric function.

Hypergeometric CS are not orthogonal (in general) since 
\be
\langle z;\alpha,\beta|z';\alpha,\beta\rangle=\frac{{}_pF_q(\alpha,\beta;\bar z z')}{[{}_pF_q(\alpha,\beta;|z|^2){}_pF_q(\alpha,\beta;|z'|^2)]^\um},\label{hypercsoverlap}
\ee
but they form an overcomplete set and close a resolution of the identity
\be
\int_{\mathbb{C}}\frac{d^2z}{\pi}{}_p\tilde{\omega}_q(|z|^2)\,|z;\alpha,\beta\rangle\langle z;\alpha,\beta| 
=\sum_{n=0}^\infty |n\rangle\langle n|=\mathbb{1},\label{resolhCS}
\ee
with a weight function ${}_p\tilde{\omega}_q$. Writing ${}_p{\omega}_q(x)={}_p\tilde{\omega}_q(x)/{}_pF_q(\alpha,\beta;x)$, with $x=|z|^2$, this 
function must be a solution of the Hausdorff moment problem  
\be
\label{mmpro}
\int_{0}^{R} {}_p{\omega}_q(x)\, x^{n}\,dx={}_p\rho_q(n), \quad x=|z|^2.
\ee
where $R=1$ or $R=\infty$. These are classical 
mathematical problems on which an extensive and mathematically oriented literature
exists, see for instance  \cite{Akhiezer,Tamarkin,Simon} and references therein. The solution 
can be obtained  by using Mellin transform techniques \cite{Klauder&PensonS,QuesneC} and has the form
\bea
 {}_p{\omega}_q(x)=\frac{\Gamma(\alpha_1)\cdots\Gamma(\alpha_p)}{\Gamma(\beta_1)\cdots\Gamma(\beta_q)} \,G_{p,q+1}^{q+1,0}\left(x\,\Bigg|\begin{array}{c}\alpha_1-1, \cdots,\alpha_p-1\\
\beta_1-1,\cdots,\beta_q-1,0\end{array}\right)\,,
\eea
where $G$  is the Meijer function \cite{Marichev,grad}. 

Note that although HCS close a resolution of the identity \eqref{resolhCS}, and therefore the $P$-function for the density operator associated with a HCS is a Dirac delta on $\mathbb{C}$ (as in the case of
canonical CS), HCS cannot be considered as classical states, i.e. except for the canonical case ($f(n)=1$), they possess nonclassical features like non-Poissonian distributions (see Sec. \ref{statprop}), squeezing or anti-bunching (see, for instance, \cite{Zelaya}).

A duality transformation was discussed by \cite{AliTavassoly} which, in the case of nonlinear $f$-CS reduces to $f\to 1/f$. For $f$ given in \eqref{fhypergeom}, 
this means $p\leftrightarrow q$ and $\alpha\leftrightarrow \beta$. Note that, given a Barut-Girardello eigenstate of $\hat{a}_f$  like \eqref{expoF}, it can also be written as 
a displaced vacuum (an exponential action)
\be
|z;\alpha,\beta\rangle={}_pF_q(\alpha,\beta;|z|^2)^{-\um} \exp\left(z f(\hat{n})^{-1} \hat{a}^\dag\right)|0\rangle,\label{expoF2}
\ee
of the dual nonlinear creation operator $\hat{a}_{1/f}^\dag=f(\hat{n})^{-1} \hat{a}^\dag$. In this sense, Barut-Girardello CSs of $\hat{a}_f$ are  ``exponential'' CSs of the dual $\hat{a}_{1/f}$ (in certain cases, when $\hat{a}_f$ and $\hat{a}_f^\dag$ close a Lie algebra, they are Gilmore-Perelomov CSs).

Canonical ($p=0=q$),  $SU(1,1)$,  Susskind-Glogower, etc, CS are recovered as particular cases of HCS. 
Let us discuss a selection of interesting and paradigmatic cases in more detail, together with their dual cases (see also Table \ref{TableHyperCS} and \ref{TableTruncHyperCS} 
for an account of the most interesting examples of HCS and some of their properties).

\subsubsection{Barut-Girardello ($p=0, q=1, \beta_1=2s$) and Perelomov  ($p=1, q=0, \alpha_1=2s$) SU(1,1) CS .}

For the case $p=0, q=1, \beta_1=2s$, we have that $f(n)=\sqrt{2s+n-1}$, and therefore $\ah_{f}=K_-$, where $K_-$ is the annihilation operator of $SU(1,1)$ algebra:
\be
K_- =\sum_{n=0}^\infty \sqrt{(n+1)(2s+n)}|n\rangle\langle n+1|
\ee
The corresponding HCS are:
\begin{equation}
 |z;\cdot,2s\rangle={}_0F_1(\cdot,2s;|z|^2)^{-1/2}\sum_{n=0}^\infty \left(n!\sqrt{\binom{2s+n-1}{n}}\right)^{-1}{z^n}|n\rangle\,,
\end{equation}
where ${}_0F_1(\cdot,2s;|z|^2)$ is related to a modified Bessel function, and they are defined on the whole complex plane. These are the original Barut-Girardello CS for $SU(1,1)$ \cite{BarutGirardello}.

The dual case is given  $p=1, q=0, \alpha_1=2s$,  $1/f(n)=1/\sqrt{2s+n-1}$, and  $\ah_{1/f}=\sum_{n=0}^\infty \sqrt{\frac{n+1}{2s+n}}|n\rangle\langle n+1|$. The associated  HCS are given by:
\begin{equation}
 |z;2s,\cdot\rangle=(1-|z|^2)^s\sum_{n=0}^\infty \sqrt{\binom{2s+n-1}{n}}{z^n}|n\rangle\,.
\end{equation}
They coincide with Perelomov's coherent states. Note that $\ah_{1/f}\not=K_-$ in this case.

\subsubsection{Susskind-Glogower ($p=1, q=0, \alpha_1=1$) and its dual ($p=0, q=1, \beta_1=1$) CS}

Let us study  the case $p=1, q=0, \alpha_1=1$ (Sudarshan harmonius or Susskind-Glogower).  For this case $f(n)=\frac{1}{\sqrt{n}}$, thus the annihilation and creation operators are:
\begin{equation}
 \ah_f=\sum_{n=0}^\infty |n\rangle\langle n+1|,\quad \ad_f=\sum_{n=0}^\infty |n+1\rangle\langle n| \,.\label{SG}
\end{equation}
In this case we recover Susskind-Glogower phase operators \cite{Susskind} $\hat{V}=\ah_f$ and $\hat{V}^\dag=\ad_f$
satisfying $ \hat{V}|n\rangle = |n-1\rangle$, $\hat{V}^\dagger|n\rangle = |n+1\rangle,\, n\in \mathbb{N}$ ($\hat V|0\rangle = 0$). 
Susskind-Glogower phase operators  satisfy $\hat{V}\hat{V}^\dagger=\mathbb{1}$ and   $\hat{V}^\dagger\hat{V}=\mathbb{1}-|0\rangle\langle 0|$. 
Therefore, these operators are not unitary, but they constitute a partial isometry. They neither close a Lie algebra, since $[\hat{V},\hat{V}^\dagger]=|0\rangle\langle 0|$. 
They were introduced in \cite{Susskind} as candidates for quantum phase operators, but there are fundamental issues with the interpretation of these operators 
(see \cite{NietoReview} for an historical review) as quantum phase 
operators. See Pegg \& Barnett \cite{Pegg1,Pegg2} for the introduction of new phase operators which solve some of these problems. 
Nonlinear $f$-CS for this case  lead to 
\begin{equation}
 |z;1,\cdot\rangle = \sqrt{1-|z|^2} \sum_{n=0}^\infty z^n |n>\,.
\end{equation}
The series is convergent for $|z|<1$, since the normalization factor is $\mathcal{N}_f(|z|)=\sum_{n=0}^\infty |z|^{2n}=\frac{1}{1-|z|^2}$ in this case. 
These coherent states are true eigenstates of the annihilation operator $\hat{V}$ in this case.

The dual case (in the sense of \cite{AliTavassoly}), $p=0, q=1, \beta_1=1$, corresponds to $f(n)={\sqrt{n}}$ and 
\begin{equation}
 |z;\cdot,1\rangle = I_0(2|z|)^{-1/2} \sum_{n=0}^\infty \frac{z^n}{n!} |n>\,,
\end{equation}
where $I_0(2|z|)$ is the modified Bessel function (see Table \ref{TableHyperCS}).

The case of CS on the circle \cite{Louisell,Newton} could also be seen as a particular case of HCS 
(namely,  $p=1, q=0, \alpha_1=1$) if we extend the summation in 
(\ref{af},\ref{nonlinearfCS}) to the whole integers (see later on appendix \ref{CScircle} for more information on this case).

\subsubsection{Inverse bosonic operator CS  ($p=2, q=0, \alpha_1=1=\alpha_2$) and Hydrogen-like spectrum 
($p=3, q=0, \alpha_1=1, \alpha_2=2=\alpha_3$)}

In \cite{Tavassoly} inverse bosonic operators are introduced as pseudo-inverses of the usual (either canonical of $f$-deformed) bosonic operators. It turns
out the the inverse of the creation operator acts as an annihilation operator, and therefore Barut-Girardello CS can be defined for it. It is shown that 
inverse bosonic operators are $f$-deformed bosonic operators with $f(n)=\frac{1}{nf(n)^*}$.

Inverse bosonic CS for canonical bosonic operators correspond to hypergeometric $f$-deformed bosonic operators with $p=2, q=0, \alpha_1=1=\alpha_2$. Coherent states in this case are ill-defined, since
the radius of convergence is $R=0$ (see Table \ref{TableHyperCS}). However, the dual case (in the sense of \cite{AliTavassoly}) is well-defined, corresponding to the case ($p=0, q=2, \beta_1=1=\beta_2$) (see Table \ref{TableHyperCS}).

Hydrogen-like spectrum $ E_n=1-\frac{1}{(n+1)^2}\,, n=0,1,\ldots\,,$ 
can also be reproduced with an $f$-deformed oscillator choosing $f(n)=\frac{\sqrt{n+2}}{n+1}$ (see \cite{Tavassoly}), corresponding to $p=2, q=1, \alpha_1=2=\alpha_2,\beta_1=3$.

\subsection{\label{hypersec2} Truncated Hypergeometric CS}

We shall not restrict ourselves to positive $\alpha$ and $\beta$, but we shall also consider negative integer values. 
In this case, the function $f(n)$ in \eqref{fhypergeom} has poles and zeros that affect the definition of  
$\ah_f$, $\ad_f$ and $|z;f\rangle$ in \eqref{af} and \eqref{nonlinearfCS}. Therefore, for analytic reasons, one must 
truncate summations in both \eqref{af} and \eqref{nonlinearfCS}. The upper limit $N$ of the summation must be chosen as the absolute value of the 
largest negative $\alpha_j$ or $\beta_j$. The normalization factor $\mathcal{N}_f$ 
is a truncation of the hypergeometric function \eqref{normfact} with absolute values for the Pochhammer symbols
\be
\mathcal{N}_f(|z|)=\sum_{n=0}^N \frac{|z|^{2n}}{{n!} |f(n)!|^2}=
\sum_{n=0}^N \frac{|(\alpha_1)_n| \cdots |(\alpha_p )_n| }{ |(\beta_1)_n| \cdots |(\beta_q)_n| }\frac{|z|^{2n}}{n!}\,.\label{normfactN}
\ee
We can still write this normalization in the form of a truncated hypergeometric function. Indeed, let us denote by $\varsigma$ the number of negative 
components of $\alpha$ and $\beta$. Then $\mathcal{N}_f(|z|)={}_pF_q(\alpha,\beta;(-1)^\varsigma|z|^2)_N$, the truncated hypergeometric function, where 
$(-1)^\varsigma$ has been introduced to compensate the negative sign of the Pochhammer symbol, namely 
$(-|\alpha_j|)_n=(-1)^n(|\alpha_j|)_n$, for $n\leq |\alpha_j|$. This truncation process implies that $|z;f\rangle$ is not an eigenstate of $\ah_f$ anymore, although 
it can still be considered an ``almost eigenstate'' for large $N$ in the following sense.

\begin{proposition}\label{quasiBarut} The distance from  $\ah_f|z;f\rangle$ to $z|z;f\rangle$ is bounded for large $N$ by
\[  \|\ah_f|z;f\rangle-z|z;f\rangle\|^2< \frac{|z|^{2(N+1)}}{(N!)^{q-p+1}},\]
which tends to zero as $N\to\infty$ for any $z\in\mathbb{C}$ as long as $q>p-1$. For $q=p-1$, $|z|<1$ is required.
\end{proposition}
\noindent \textbf{Proof}: A direct computation gives 
\[  \|\ah_f|z;f\rangle-z|z;f\rangle\|^2= \mathcal{N}_f^{-1}(|z|)\frac{|z|^{2(N+1)}}{N!(f(N)!)^2}.\]
The large $N$ behavior of $f(N)$ in \eqref{fhypergeom} gives $f(N)^2\sim N^{q-p+1}$, so that $\frac{|z|^{2N}}{N!(f(N)!)^2}$ is smaller and smaller for $q>p-1$. 
Since $\mathcal{N}_f(|z|)=1+\dots+\frac{|z|^{2N}}{N!(f(N)!)^2}$ (all addends positive), we can say that  $\mathcal{N}_f^{-1}(|z|)\frac{|z|^{2(N+1)}}{N!(f(N)!)^2}<\frac{|z|^{2(N+1)}}{N!(f(N)!)^2}$. 
The rest is a consequence of the Stirling formula.
$\square$ 

Let us analyze the  particular interesting case of $SU(2)$  Barut-Girardello and its dual (Perelomov). 
Let us consider $p=0, q=1, \beta_1=-2s$. In this case $N=2s$, and the dimension of the Hilbert space is $2s+1$.  
The normalization factor \eqref{normfactN} is the  hypergeometric function ${}_0F_1(\cdot,-2s;-|z|^2)_{2s}$ truncated to the first $2s+1$ addends. The nonlinear 
annihilation operator $\ah_f=i\sum_{n=0}^{2s} \sqrt{(n+1)(2s-n)}|n\rangle\langle n+1|$ coincides with the angular momentum ladder operator $iJ_-$. As we mentioned before, the truncated HCS 
 \be |z;\cdot,-2s\rangle={}_0F_1(\cdot,-2s;-|z|^2)_{2s}^{-1/2}\sum_{n=0}^{2s}\binom{2s}{n}^{-1/2} \frac{(iz)^n}{n!}|n\rangle\,,
 \ee
is not an eigenstate  of $J_-$, but it is ``almost'' an eigenstate for large spin $s$ in the sense of the Proposition \ref{quasiBarut}. In this sense, we can denote them as  ``almost'' Barut-Girardello CS for $SU(2)$ 
 and they have been previously introduced by \cite{PopovspincsBG}. In   \cite{RoyRoy} and \cite{Angelova}  they relate these CS with the Morse potential using Gazeau--Klauder coherent states (see also \cite{Cotfas}), 
 although the sum is not up to $N=2s$ but up to $\nu=s-1$ for integer $s$, or $\nu=\lfloor\lceil(2s-1)\rceil/2\rfloor$ for non integer $s$ (in terms of standard floor and ceiling functions), 
 in this case expressing the binomial coefficients and the factorial  in terms of the corresponding 
 Gamma functions;  the rest of the terms are not present in the spectrum of the Morse potential, as they would correspond to non-normalizable or anti-bound states \cite{Inconsistencies}. 
 The dual case $p=0, q=1, \alpha_1=-2s$ is related to  Perelomov spin-$s$ $SU(2)$ CS. Indeed, we recognize the binomial structure of 
$\ah_f=-i\sum_{n=0}^{2s} \sqrt{(n+1)/(2s-n)}|n\rangle\langle n+1|$. CS
\be |z;-2s,\cdot\rangle={}_1F_0(-2s,\cdot;-|z|^2)_{2s}^{-1/2}\sum_{n=0}^{2s}
 \binom{2s}{n}^{1/2} (iz)^n|n\rangle\,,
 \ee
with  ${}_1F_0(-2s,\cdot;-|z|^2)_{2s}=(1+|z|^2)^{2s}$  the usual Bergman kernel for $SU(2)$.

\section{$k$-th order nonlinear harmonic generation and  hypergeometric Schr\"odinger kittens}\label{kcatsec}

In this section we shall compute the eigenstates of $\ah_f^k$ for $k>1$ and positive $\alpha_j$ and $\beta_j$, which appear in the $k$-th 
order harmonic generation of nonlinear $f$-CS. We already know that HCS of equation \eqref{hypercs} are eigenstates of $\ah_f^k$ for $k=1$. For 
negative  $\alpha_j$ and $\beta_j$, the truncation process affects the definition of Schr\"odinger kittens (they are no more eigenstates of $\ah_f^k$), but 
an alternative definition can still be considered (see Section \ref{hyperlikecats}).

\subsection{Ortonormalized eigenstates of $\ah_f^k$ and $k$-hypercats}

Now we are interested in the eigenstates of $\ah_f^k$ for $k>1$. The eigenvalues of $\ah_f^k$ are of the form $z^k$, and the corresponding eigenspaces are $k$-fold degenerated, spanned by the HCS 
$\{|z e^{i\frac{2\pi j}{k}};\alpha,\beta\rangle\,,\,\,j=0,1\,\ldots,k-1\}$, i. e. HCS with $z$ given by the $k$-th roots of $z^k$, and therefore are obtained  from $|z ;\alpha,\beta\rangle$  
by rotating $z$ counter-clock-wise successively  by an angle $\frac{2\pi}{k}$. 
Using eqn. \eqref{hypercsoverlap}, the overlap between these states is given by:
\begin{equation}
 G_{jl}=\langle z e^{i\frac{2\pi j}{k}};\alpha,\beta|z e^{i\frac{2\pi l}{k}};\alpha,\beta\rangle = 
 \frac{{}_pF_q(\alpha,\beta;|z|^2e^{ i\frac{2\pi (l-j)}{k}} )}{{}_pF_q(\alpha,\beta;|z|^2)} \equiv C_{l-j}.\label{gram}
\end{equation}
This means that these vectors are non-orthogonal. We have the following result:

\begin{proposition}
 An orthonormalized basis for the eigenspace of eigenstates of $\ah_f^k$ is given by:
 \bea
|z;\alpha,\beta;k,j\rangle =  \frac{1}{k}\left(\frac{{}_pF_q(\alpha,\beta;|z|^2)}{{}_p^kF_q^j(\alpha,\beta;|z|^2)}\right)^{1/2}
\sum_{l=0}^{k-1}e^{-i\frac{2\pi  jl}{k}}|ze^{i\frac{2\pi  l}{k} };\alpha,\beta\rangle,\quad j=0, 1, \cdots k-1.\label{discretecircle} 
\eea
This basis can be rewritten in the usual form:
\bea
\label{reaais}
|z;\alpha,\beta;k,j\rangle={}_p^kF_q^j(\alpha,\beta;|z|^2)^{-\um} \sum_{n=0}^\infty\frac{z^{nk+j}}{\sqrt{{}_p\rho_q(nk+j)}} |nk+j\rangle,\quad j=0,1,\dots k-1,\label{khypercats}
\eea
where ${}_p^kF_q^j(\alpha,\beta;|z|^2)$ are normalization coefficients (see appendix \ref{apA} for its expression in terms of hypergeometric functions).
\end{proposition}

\noindent \textbf{Proof}: The  corresponding Gram matrix  $G$ in \eqref{gram} for the non-orthogonal set $\{|z e^{i\frac{2\pi j}{k}};\alpha,\beta\rangle\,,\,\,j=0,1\,\ldots,k-1\}$ has a circulant structure 
(see \cite{samplingsphere} and references therein), since it depends only on the difference $l-j$. The Gram orthonormalization process is then performed by the Discrete Fourier Transform (DFT) ${\cal F}_k$:
\bea
\label{gramortho}
|z;\alpha,\beta;k,j\rangle= \lambda_j^{-\um} \frac{1}{\sqrt{k}} \sum_{l=0}^{k-1}e^{-i\frac{2\pi  jl}{k}}|ze^{i\frac{2\pi  l}{k} };\alpha,\beta\rangle ,\quad j=0, 1, \cdots k-1,
\eea
where $\vec{\lambda}= \sqrt{k}{\cal F}_k \vec{C}$ are the eigenvalues of the Gram matrix, which are given by:
\bea
\lambda_j&=& \sum_{l=0}^{k-1} e^{-i\frac{2\pi  jl}{k}} C_l =\sum_{l=0}^{k-1} e^{-i\frac{2\pi  jl}{k}} \frac{{}_pF_q(\alpha,\beta;|z|^2e^{ i\frac{2\pi l}{k}} )}{{}_pF_q(\alpha,\beta;|z|^2)} 
=  \frac{1}{{}_pF_q(\alpha,\beta;|z|^2)}   \sum_{n=0}^\infty \frac{|z|^{2n}}{{}_p\rho_q(n)} \sum_{l=0}^{k-1} e^{-i\frac{2\pi  jl}{k}}  e^{ i\frac{2\pi n l}{k}} \nonumber \\
& = & \frac{k}{{}_pF_q(\alpha,\beta;|z|^2)} \sum_{\nu=0}^\infty \frac{|z|^{2(\nu k+j)}}{ {}_p\rho_q(\nu k+j)} \equiv  k\, \frac{{}_p^kF_q^j(\alpha,\beta;|z|^2)}{{}_pF_q(\alpha,\beta;|z|^2)},\quad j=0, 1, \cdots k-1, \label{eigenkats}
\eea
where we have used the orthogonality  relation
\begin{equation}
 \sum_{l=0}^{k-1} e^{2\pi i(n-j)l/k}=k \delta_{n,j \mathrm{mod}\, k}\,.\label{orthogonality}
\end{equation}
With this, the orthonormalized basis \eqref{gramortho} is written as \eqref{discretecircle}. Moreover, using an argument similar to that used in \eqref{eigenkats}, the orthonormalized basis also adopts the usual expression \eqref{khypercats}. 
The proof does not depend on the fact that we are working with hypergeometric coherent states, and a similar result holds for 
an arbitrary $f$ (see also \cite{LiuJPA32} for another derivation for arbitrary $f$).$\square$
 
Note that the states \eqref{discretecircle} have the structure of an equally-weighted superposition of phase-shifted HCS. 
Therefore, we shall call these states  ``hypergeometric Schr\"odinger kittens'' or ``$k$-hypercats'' for short, in the sense that 
they are  quantum superpositions of $k$ semi-classical (macroscopic) CS  equally distributed on the circle of radius $r=|z|$. The overlap 
$\langle ze^{2\pi i l/k };\alpha,\beta|ze^{2\pi i m/k };\alpha,\beta\rangle$, $l\not=m$,  will be negligible as long as $2\pi |z|/k$ is large enough (see Sec. \ref{kerrsec} for a quantitative analysis of this distance). The case $k=2$ is nothing but the even 
and odd nonlinear CS discussed in \cite{Mancini} for the case of hypergeometric functions $f$. The particular case $p=q=0$  (canonical CS) was extensively studied long ago by 
Jinzuo {\it et al} \cite{PRA44,PRA46} and, as we have said, the general $f$ case has been studied by \cite{LiuJPA32,JPB35}, among others. Here we shall further study the properties of these 
interesting states for the case of hypergeometric $f$. 

The $k$-hypercats \eqref{khypercats} are normalized and  orthogonal in the index $j$ (but they are not orthogonal in the index $z$), and their overlap is given by

\be
\langle z;\alpha,\beta;k,j|z';\alpha,\beta;k,l\rangle=\frac{{}_p^kF_q^j(\alpha,\beta;\bar z z')}{[{}_p^kF_q^j(\alpha,\beta;|z|^2){}_p^kF_q^l(\alpha,\beta;|z'|^2)]^\um}\delta_{jl}.
\ee
As for HCS, $k$-hypercats form an overcomplete set and close the resolution of the identity
\be
\label{pmo}
\int_{\mathbb{C}}\frac{d^2z}{\pi}\sum_{j=0}^{k-1} \,{}_p^k\omega_q^j(|z|^2)\,|z;\alpha,\beta;k,j\rangle\langle z;\alpha,\beta;k,j| =\mathbb{1},
\ee
where ${}_p^k \omega_q^j(|z|^2)={}_p{\omega}_q(|z|^ 2){}_p^kF_q^j(\alpha,\beta;|z|^2)$. This closure relation can be easily obtained from \eqref{resolhCS}. 

We shall study Canonical and Perelomov $SU(1,1)$ Schr\"odinger kittens as  representative cases of coherent states on $\mathbb{C}$ and on the unit disk, respectively. For canonical $k$-cats, the 
normalization factor in \eqref{khypercats} adopts the following form (see \cite{PRA44}) 
\be
{}_0^kF_0^j(\cdot,\cdot;|z|^2)=\sum_{l=0}^{k-1} e^{2\pi il(k-j)/k}\exp(|z|^2 e^{2\pi il/k}).
\ee
For example, for $k=2$ we have ${}_0^2F_0^0(\cdot,\cdot;|z|^2)=\cosh(|z|^2)$ and ${}_0^2F_0^1(\cdot,\cdot;|z|^2)=\sinh(|z|^2)$. 
For Perelomov $SU(1, 1)$ CS ($p=1,\;q=0, \alpha_1=2s$) we have 
\be
{}_1^kF_0^j(2s,\cdot;|z|^2)= \frac{(2s)_j |z|^{2j}  }{  \,j!}{}_{ k+1}F_{ k}\left(
\begin{array}{c}
 \frac{2s+j}{k}, \frac{2s+j+1}{k},    \cdots, \frac{2s+j+k-1}{k}, 1\\
  \frac{j+1}{k},  \frac{j+2}{k} , \cdots,  \frac{j+k}{k} 
\end{array}\Bigg|{|z|^{2k}}
\right).
\ee
In particular, for $k=2$, we have \[{}_1^2F_0^0(2s,\cdot;|z|^2)={}_{2}F_{1}\left(
\begin{array}{c}
s,s+\frac{1}{2}\\
\frac{1}{2}\end{array}\Big| |z|^4 
\right),\quad {}_1^2F_0^1(2s,\cdot;|z|^2)={}_{2}F_{1}\left(
\begin{array}{c}
s+\frac{1}{2}, s+1\\
\frac{3}{2}\end{array}\Big|  |z|^4 
\right).\]
The special case $s=1/2$ corresponds to Susskind-Glogower $k$-cats, which normalization factor adopts the simple form:
\be
{}_1^kF_0^j(1,\cdot;|z|^2)=\frac{|z|^{2j}}{1-|z|^{2k}}.
\ee

\subsection{Statistical properties}\label{statprop}

Let us now study the photon number statistics of $k$-hypercats. The probability of detecting $m$ quanta in a $k$-hypercat $|z;\alpha,\beta;k,l\rangle$ is given by the photon number distribution
\be
{}_p^kP_q^j(\alpha,\beta;m;|z|^2)=|\langle m|z;\alpha,\beta;k,j\rangle|^2={}_p^kF_q^j(\alpha,\beta;|z|^2)^{-1}\frac{|z|^{2m}}{{{}_p\rho_q(m)}}\delta_{j,m\, \mathrm{mod}\, k}.
\ee
We shall study, in particular, the two special cases
\begin{enumerate}
\item For canonical ($p=0=q$) $k$-cats  we have ${}_0^kP_0^j(\cdot,\cdot;m;|z|^2)=\frac{|z|^{2m}\delta_{j,m\, \mathrm{mod}\, k}}{{}_0^kF_0^j(\cdot,\cdot;|z|^2)m!}$
% \be
% {}_0^2P_0^0(\cdot,\cdot;m;|z|^2)=\frac{|z|^{2m}\;\delta_{m,0\, \mathrm{mod} \;2}}{m!\;\cosh(|z|^2)},\quad {}_0^2P_0^1(\cdot,\cdot;m;|z|^2)=\frac{|z|^{2m} \delta_{m,1\, \mathrm{mod}\; 2}}{m!\; \sinh(|z|^2)}.
% \ee
\item For Perelomov SU(1,1) ($p=1, q=0, \alpha_1=2s$) $k$-cats we have   
${}_1^kP_0^j(2s,\cdot;m;|z|^2)={}_1^kF_0^j(2s,\cdot;|z|^2)^{-1}\frac{(2s)_m|z|^{2m}}{m!}\delta_{j,m\, \mathrm{mod}\, k}$.

\end{enumerate}

Note that, for the special case $s=1/2$ (Susskind-Glogower) the probability ${}_1^kP_0^j(1,\cdot,m,|z|^2)=(1-|z|^{2k})\delta_{j,m\, \mathrm{mod}\, k}$ does not depend on $j$.

%\textbf{Make plots of ${}_p^kP_q^j(\alpha,\beta;m;|z|^2)$}
\begin{figure}[h]
\begin{center}
(a)\includegraphics[width=7cm]{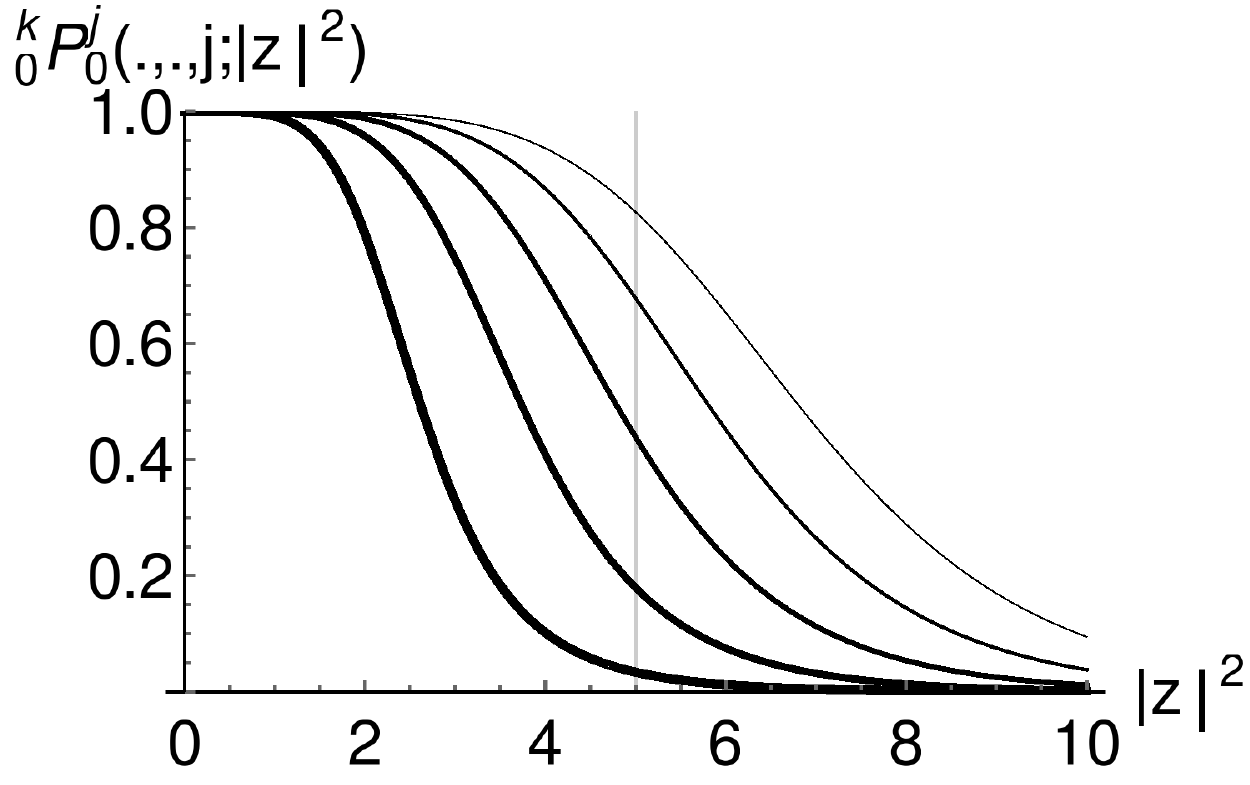}(b)\includegraphics[width=7cm]{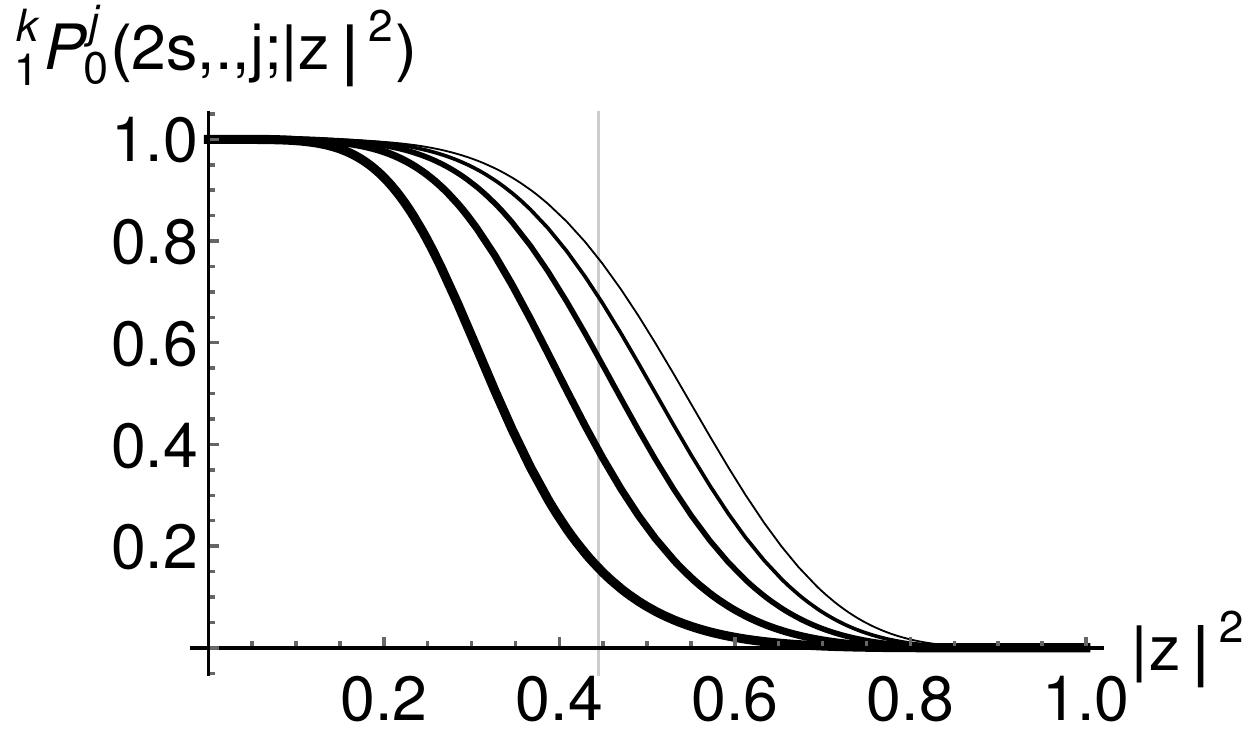}\\
(c)\includegraphics[width=7cm]{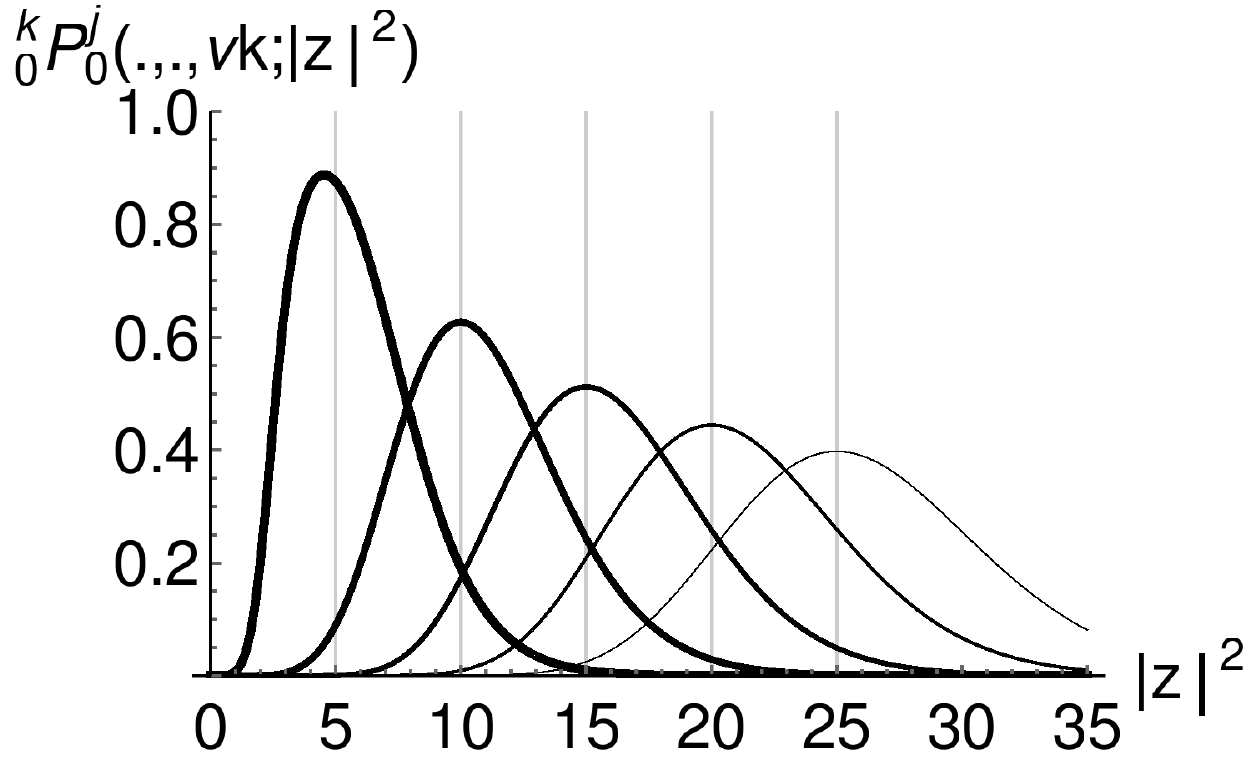}(d)\includegraphics[width=7cm]{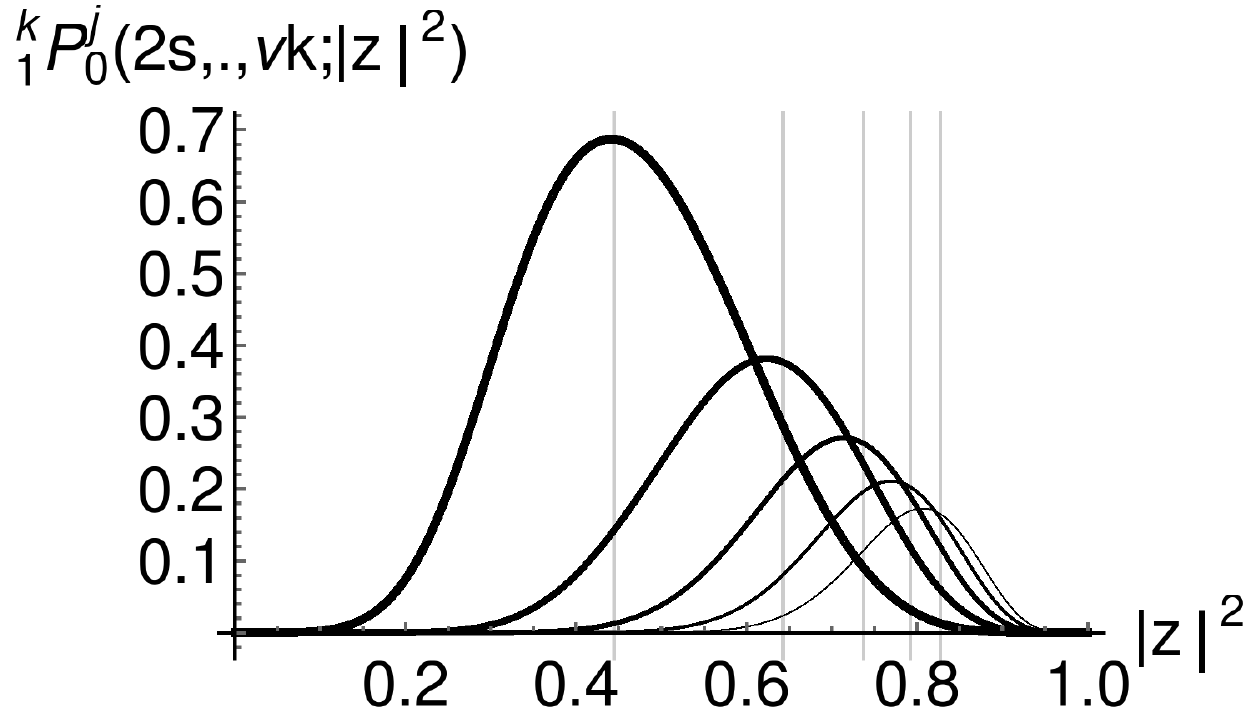}
\end{center}
\caption{(Above) Probability ${}_p^kP_q^j(\alpha,\beta;m;|z|^2)$ for: (a) Canonical CS (left, $p=0=q$),  and 
(b) Perelomov $SU(1,1)$ (right, $p=1, q=0, \alpha_1=2s=6$), for $k=5$ and $m=j=0,1,2,3,4$ (from thickest to thinest). The critical values of $|z|^2$ are marked with a vertical gridline for 
canonical,  $z_c^2=k=5$, and Perelomov $SU(1,1)$,  $z_c^2=(k-1)/(k+2s-2)=4/9$ CS.\\
(Below) Probability ${}_p^kP_q^j(\alpha,\beta;m;|z|^2)$ for: (c) Canonical CS (left)  and 
(d) Perelomov $SU(1,1)$ (right), for $k=5$ and $m=\nu k$ with $\nu=1,2,3,4,5$ (from thickest to thinest). The values of $|z|^2$ where the probabilities 
have maxima  are marked with vertical gridlines for the canonical case, approximately at $|z_{\rm max}|^2=\nu k$, and Perelomov $SU(1,1)$, 
approximately at $z_{\rm max}^2=(\nu k-1)/(\nu k+2s-2)$,  CS.}\label{Probfig}
\end{figure}
The plots in Figure \ref{Probfig} (Above) indicate that ${}_p^kP_q^j(\alpha,\beta;j;|z|^2)$ is close to a step-function 
$\Theta(|z|^2-z_c^2)=\left\{\begin{array}{lcl} 1 & \mathrm{if} & |z|<z_c\\ 0 & \mathrm{if} & |z|\geq z_c  \end{array}\right.$ for some critical value $z_c$ of $|z|$. That is, for  $|z|\ll z_c$, 
the $k$-hypercat behaves as a number state $|z;\alpha,\beta;k,j\rangle\simeq |j\rangle$. We can find a representative 
value of $z_c$ for each $k$ as follows. Consider the projector $\Pi_k=\sum_{n=0}^{k-1}|n\rangle\langle n|$ onto the subspace generated by the first $k$ number states. Take the 
average $\Pi_k(\alpha,\beta,|z|^2)=\langle \alpha,\beta;z|\Pi_k|\alpha,\beta;z\rangle$ (the operator $\Pi_k$ symbol). The value of $z_c$ can be obtained as a solution to the saddle point 
equation ${d^2 \Pi_k(\alpha,\beta,x) }/{dx^2}=0$. This gives $z_c$ as a function of $(k,\alpha,\beta)$. For example, for canonical and Perelomov SU(1,1) cases, we can explicitly compute this critical value, 
which results in $z_c^2=k$ and $z_c^2=(k-1)/(k+2s-2)$, respectively \cite{samplingHW,samplingH}. For the case of Perelomov SU(1,1), this step-function behavior is sharper and sharper for higher values of 
$k$ and $s$.  

The plots in Figure \ref{Probfig} (Below) indicate that the probabilities ${}_p^kP_q^j(\alpha,\beta;\nu k;|z|^2)$, with $\nu=1,2,\ldots$ decrease to zero when $\nu$ grows, having maxima approximately at $|z_{\rm max}|^2=\nu k$ (the larger $\nu$ the better approximation) for Canonical CS and approximately at $z_{\rm max}^2=(\nu k-1)/(\nu k+2s-2)$ (the smaller $\nu$, the better approximation) for Perelomov $SU(1,1)$ CS.

This behavior indicates that the $k$-hypercat $|z;\alpha,\beta;k,j\rangle$ is a good approximation of the number state $|j\rangle$ for $|z|\leq z_c$. Let us investigate this 
fact more closely. For it, let us compute the mean number of photons in a $k$-hypercat
\be
\langle \hat{n}\rangle(\alpha,\beta;k,j;|z|^2)={}_p^kF_q^j(\alpha,\beta;|z|^2)^{-1} \sum_{n=0}^\infty\frac{(nk+j)|z|^{2(nk+j)}}{ {}_p\rho_q(nk+j)}=
|z|^2\;\frac{{}_p^k{F_q}^{j+k-1}(\alpha,\beta;|z|^2) }{{ }_p^kF_q^j(\alpha,\beta;|z|^2)}
\ee
and the photon number standard deviation                                                                                                  
\be
\sigma_{\hat{n}}(\alpha,\beta;k,j;|z|^2)=\sqrt{\langle\hat{n}^2\rangle(\alpha,\beta;k,j;|z|^2)-\langle\hat{n}\rangle(\alpha,\beta;k,j;|z|^2)^2}.
\ee
By using the fact that, $\hat{n}^2=\hat{a}^{\dag 2}\ah^2+\hat{n}$ and 
\be
 \langle \hat{a}^{\dag 2}\ah^2\rangle(\alpha,\beta;k,j;|z|^2)=|z|^4\;\;{}_p^kF_q^j(\alpha,\beta;|z|^2)^{-1}\frac{d^2}{d^2|z|^2} {}_p^kF_q^j(\alpha,\beta;|z|^2) 
=|z|^4\;\frac{{}_p^k{F_q}^{j+2k-2}(\alpha,\beta;|z|^2) }{{ }_p^kF_q^j(\alpha,\beta;|z|^2)}
\ee
we arrive to 
\be
\sigma_{\hat{n}}(\alpha,\beta;k,j;|z|^2)=\sqrt{|z|^4\;\frac{{}_p^k{F_q}^{j+2k-2}(\alpha,\beta;|z|^2) }{{ }_p^kF_q^j(\alpha,\beta;|z|^2)}+|z|^2\;\frac{{}_p^k{F_q}^{j+k-1}(\alpha,\beta;|z|^2) }{{ }_p^kF_q^j(\alpha,\beta;|z|^2)}-\left[ |z|^2\;\frac{{}_p^k{F_q}^{j+k-1}(\alpha,\beta;|z|^2) }{{ }_p^kF_q^j(\alpha,\beta;|z|^2)}\right]^2}.
\ee

\begin{figure}[h]
\begin{center}
(a)\includegraphics[width=8cm]{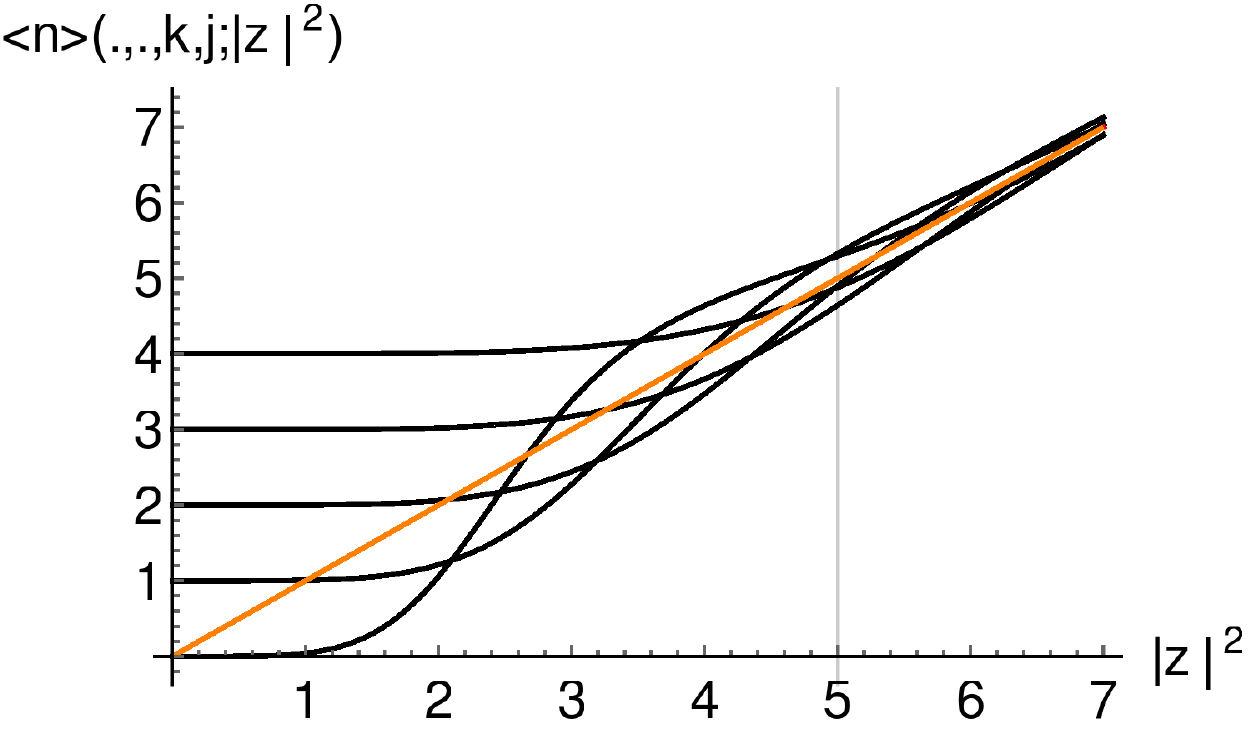}(b)\includegraphics[width=8cm]{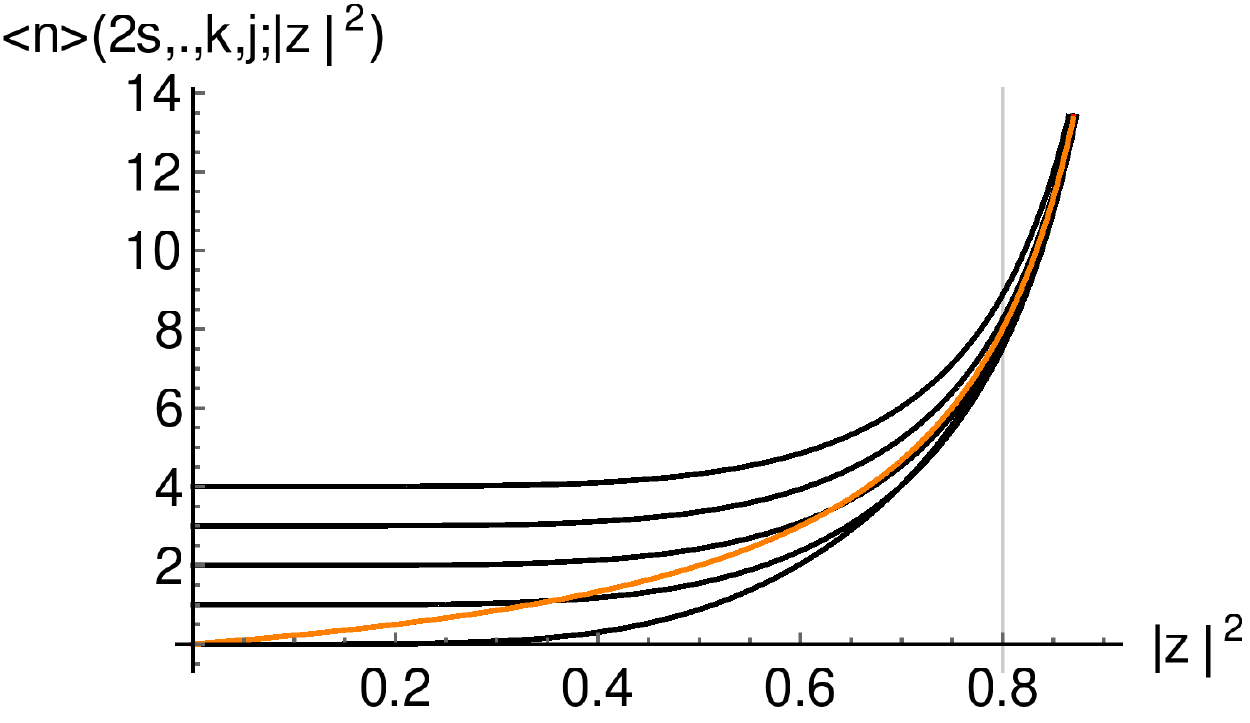}
\end{center}
\caption{Mean number $\langle \hat{n}\rangle(\alpha,\beta;k,j;|z|^2)$ of photons  for: (a) Canonical CS (left, $p=0=q$),  and 
(b) Perelomov $SU(1,1)$ (right, $p=1, q=0, \alpha_1=2s=2$), for $k=5$ and $j=0,1,2,3,4$. In orange we plot the $k=1$ case, for which $\langle \hat{n}\rangle(\cdot,\cdot;1,0;|z|^2)=|z|^2$ and 
$\langle \hat{n}\rangle(2s,\cdot;1,0;|z|^2)=\sqrt{\frac{2s x}{(x-1)^2}}$. Critical values  $z_c^2=5$ and $z_c^2=4/5$ are marked with vertical grid lines.}
\label{Probfig2}
\end{figure}

\begin{figure}[h]
\begin{center}
(a)\includegraphics[width=8cm]{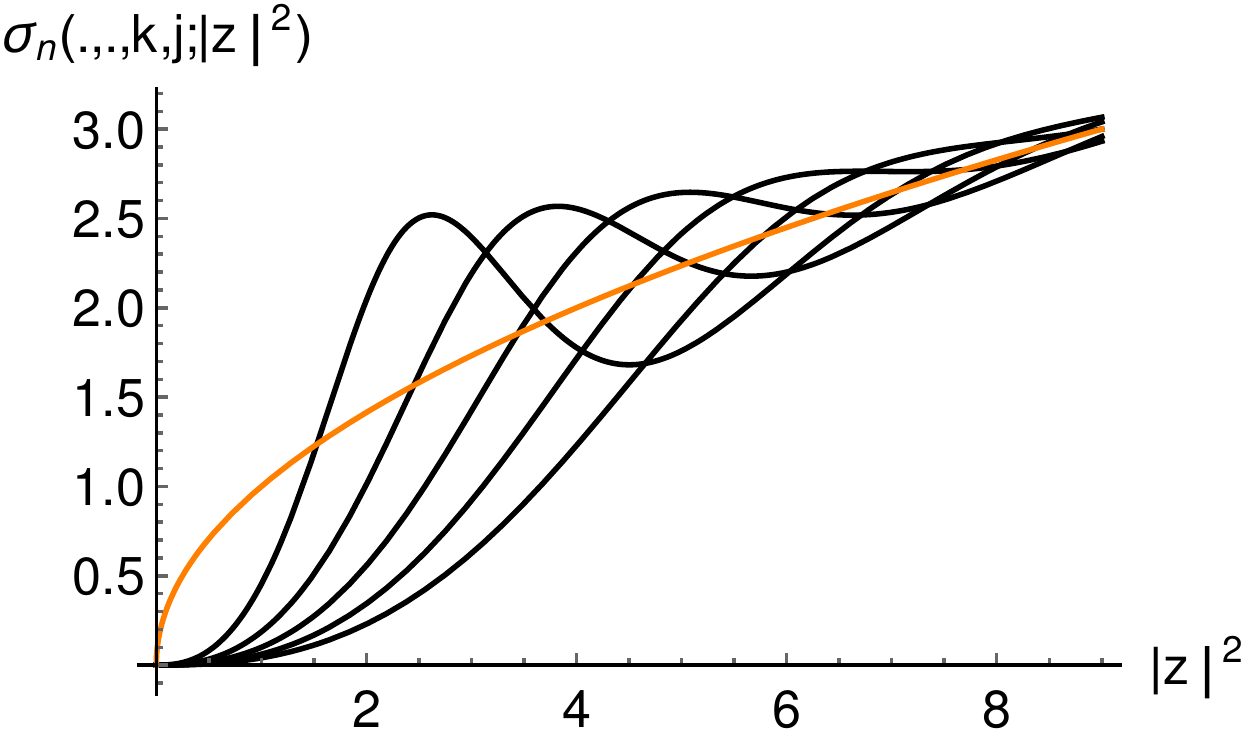}(b)\includegraphics[width=8cm]{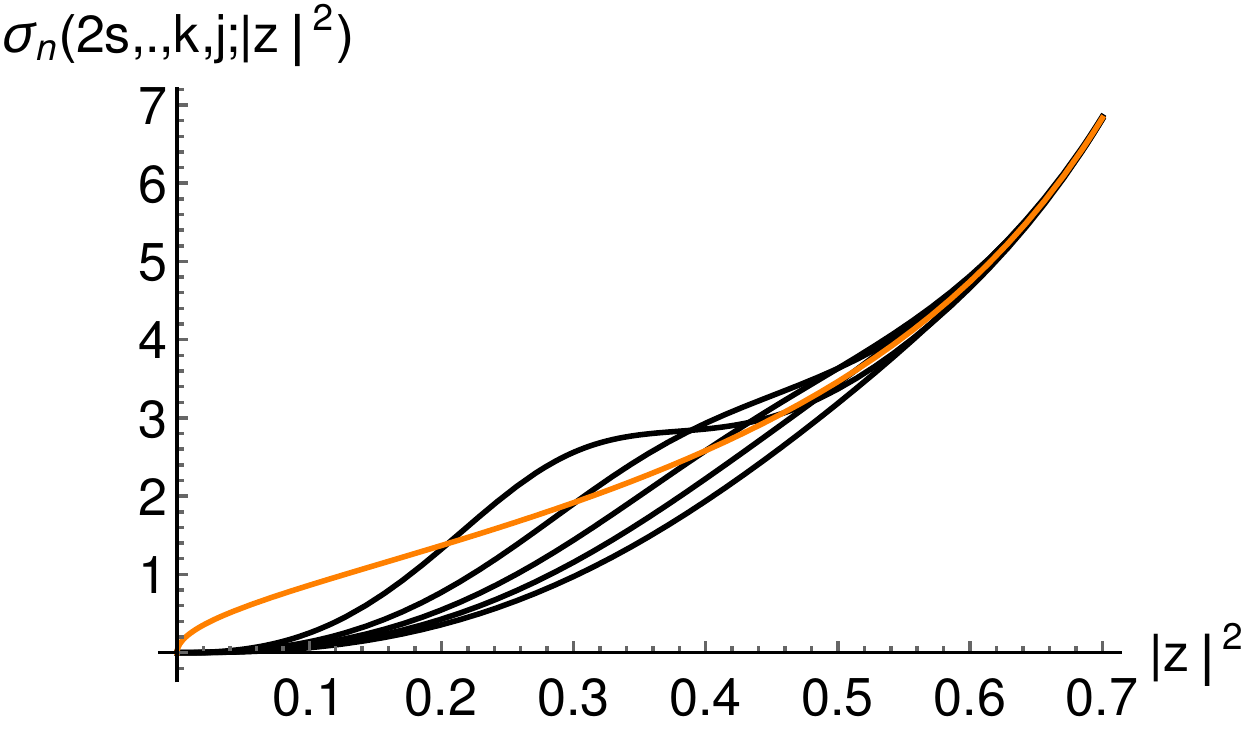}
\end{center}
\caption{Photon number standard deviation $\sigma_{\hat{n}}(\alpha,\beta;k,j;|z|^2)$  for: (a) Canonical CS (left, $p=0=q$),  and 
(b) Perelomov $SU(1,1)$ (right, $p=1, q=0, \alpha_1=2s=6$), for $k=5$ and $j=0,1,2,3,4$. In orange we plot the $k=1$ case, for which $\sigma_{\hat{n}}(\cdot,\cdot;1,0;|z|^2)=|z|$ and 
$\sigma_{\hat{n}}(2s,\cdot;1,0;|z|^2)=2s|z|^2/(1-|z|^2)$. %Critical values  $z_c^2=5$ and $z_c^2=4/5$ are marked with vertical grid lines.
}
\label{Probfig3}
\end{figure}

In Figure \ref{Probfig2} we plot the mean number of photons $\langle \hat{n}\rangle(\alpha,\beta;k,j;|z|^2)$ in a $k$-hypercat $|z;\alpha,\beta;k,j\rangle$ for canonical and 
Perelomov SU(1,1) cases, as a function of $|z|^2$. We see that, in both cases, the $k$-hypercat $|z;\alpha,\beta;k,j\rangle$ interpolates between the number state $|j\rangle$, for $|z|^2\ll z_c^2$, 
and the CS $|z;\alpha,\beta\rangle$ (the $k=1$ case), for $|z|^2\gg z_c^2$ (close to $|z|=1$ for the SU(1,1) case). Both regions (let us call them: ``number and coherent'' regions) 
are separated by a critical value $z_c$ of $|z|$, which depends on $k$ and on $\alpha$ and $\beta$. Although we have restricted ourselves to canonical and Perelomov SU(1,1) cases, the general 
behavior shown in Figure \ref{Probfig2} is representative for other values of $\alpha$ and $\beta$. In Figure \ref{Probfig3} we plot the corresponding 
photon number standard deviation $\sigma_{\hat{n}}(\alpha,\beta;k,j;|z|^2)$ and we see that whereas it is small in the number region, it grows and approaches the $\sigma_{\hat{n}}(\alpha,\beta)$ 
(the $k=1$ CS case) in the CS region. This shows again that $|z;\alpha,\beta;k,j\rangle\simeq |j\rangle$, in the number region, whereas $|z;\alpha,\beta;k,j\rangle\simeq |z;\alpha,\beta\rangle$ 
in the CS region.

\begin{figure}[h]
\begin{center}
\includegraphics[width=8cm]{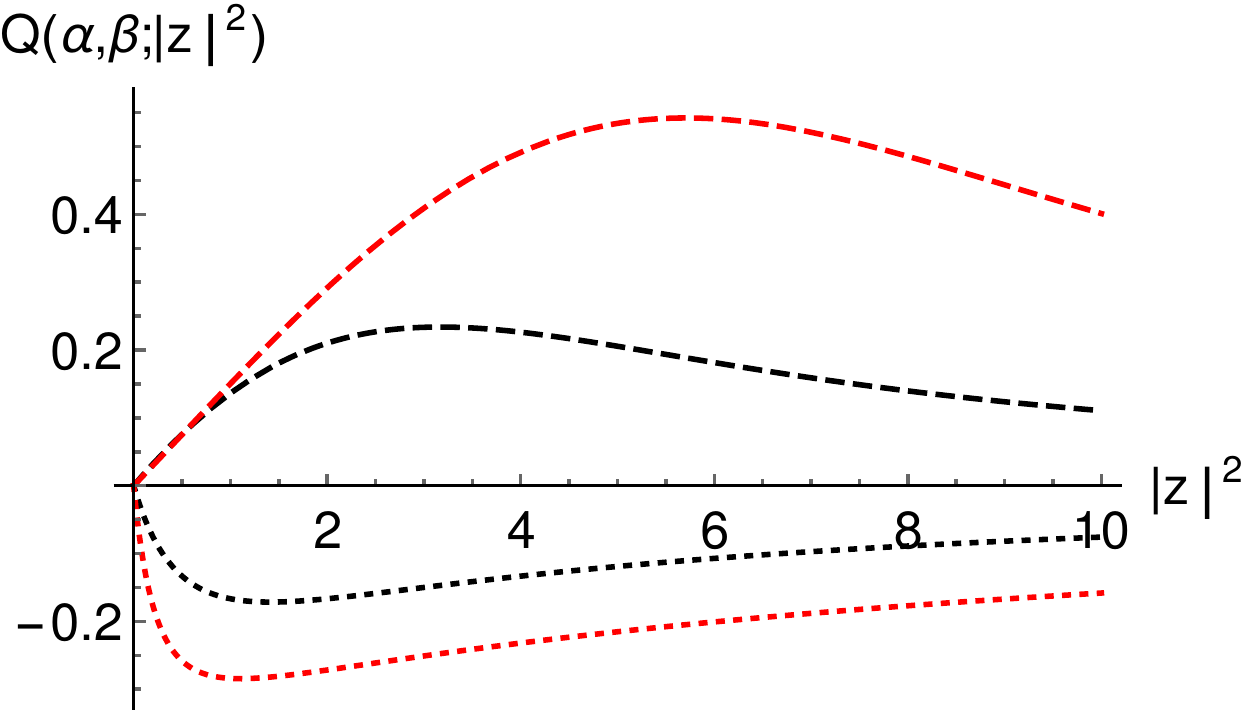}
\end{center}
\caption{Mandel parameter $\mathcal{Q}(\alpha,\beta,|z|^2)$ for confluent ($p=1=q$) HCS: 1) super-Poissonian $\alpha=1,\beta=2$ (dashed black) and $\alpha=1,\beta=4$ (dashed red) and 
2) sub-Poissonian $\alpha=2,\beta=1$ (dotted black) and $\alpha=4,\beta=1$ (dotted red).}\label{mandelab}
\end{figure}

For completeness, let us study deviations from Poissonian distributions for HCSs and its  $k$-hypercats. 
Since for canonical CS the variance $\sigma_{\hat{n}}$ of the number operator  is equal to
its average, deviations from Poisson distribution can be measured with the Mandel parameter,  defined by the quantity  
\be 
\label{sama:mand}
 \mathcal{Q}=\frac{\sigma_{\hat{n}}^2-\langle \hat{n}\rangle}{\langle \hat{n}\rangle}=\mathcal F - 1,
\ee
where $\mathcal F=\langle { \hat{n}}^2\rangle /\langle \hat{n} \rangle$ is the Fano factor \cite{bajer-miranowicz}.  For 
$\mathcal F < 1 \,( \mathcal{Q} < 0)$, the emitted light 
  is referred to as  sub-Poissonian,  for  $\mathcal F = 1,  (\mathcal{Q} = 0)$ it corresponds to  the Poisson distribution,  while 
 for $\mathcal F > 1, ( \mathcal{Q} > 0)$ it corresponds to  
 super-Poissonian \cite{mandel1,zhangetal,jpjpg}.  Let us consider for example the family of confluent ($p=1=q$) HCS 
$|z;\alpha,\beta\rangle={}_1F_1(\alpha,\beta;|z|^2)^{-\um}{}_1F_1(\alpha,\beta;z \ad)|0\rangle$. Canonical CS are a particular member of this family with $\alpha=\beta$. In Figure 
\ref{mandelab} we represent the Mandel parameter $\mathcal{Q}(\alpha,\beta,x)$ for confluent hypergeometric CS as a function of $x=|z|^2$ and several values of $\alpha, \beta$. 
We see that for $\alpha<\beta$ the corresponding distribution is super-Poissonian $(\mathcal{Q} > 0)$, whereas for $\alpha>\beta$ it is sub-Poissonian $(\mathcal{Q} < 0)$, 
the Poissonian case $(\mathcal{Q} =0)$ being $\alpha=\beta$.

\begin{figure}[h]
\begin{center}
(a)\includegraphics[width=5cm]{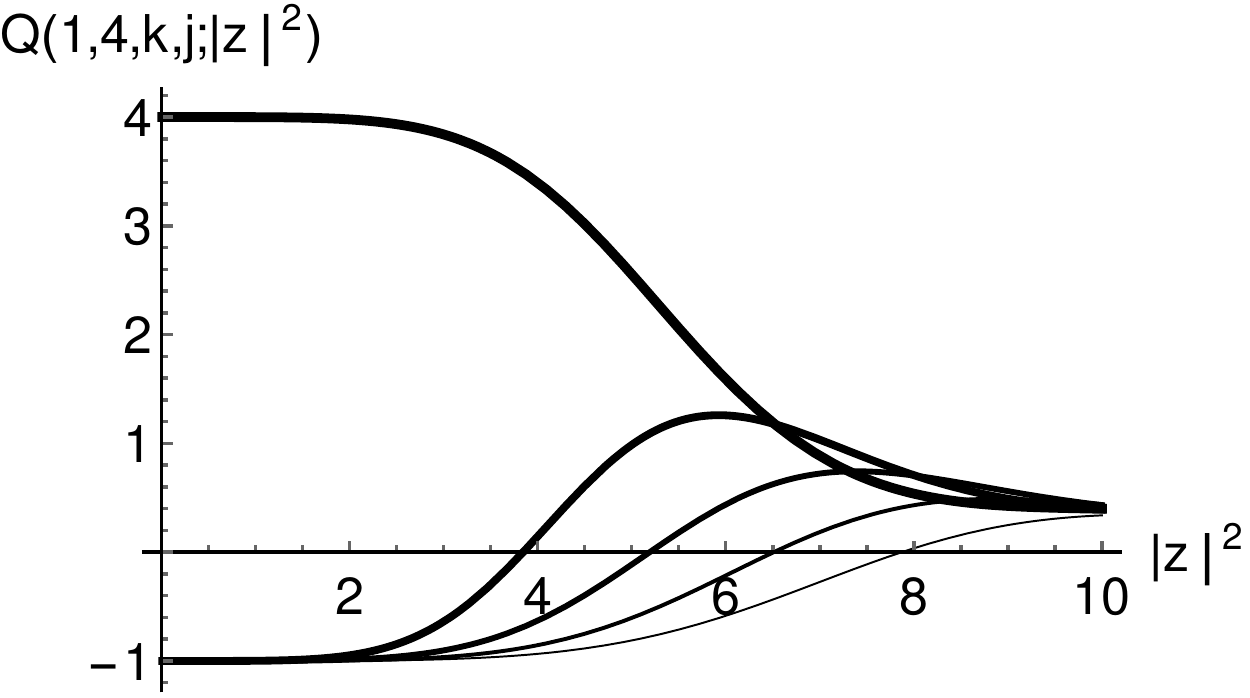}(b)\includegraphics[width=5cm]{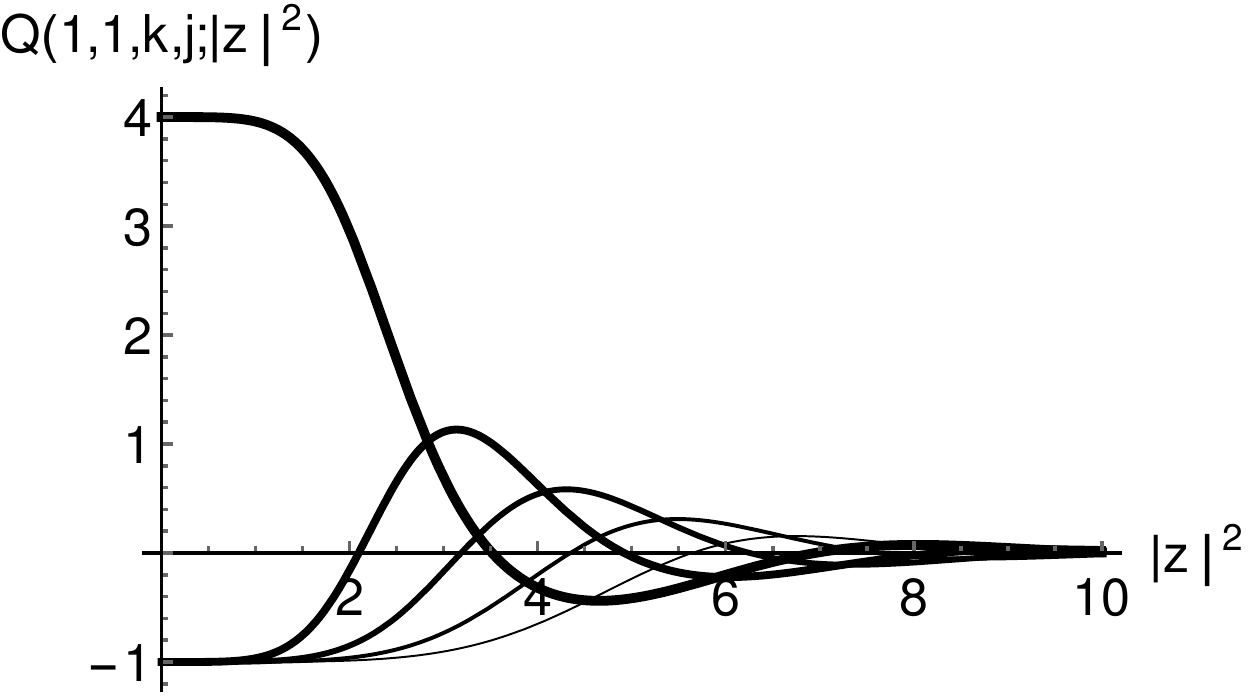}(c)\includegraphics[width=5cm]{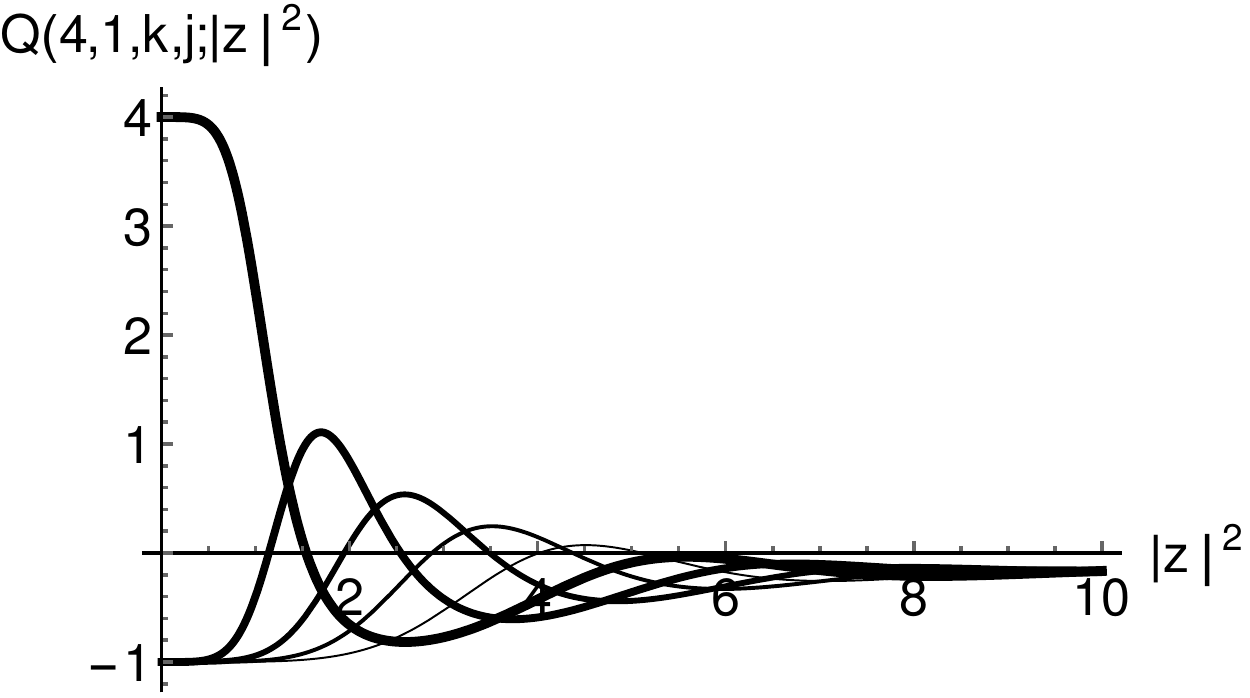}
\end{center}
\caption{Mandel parameter $\mathcal{Q}(\alpha,\beta;k,j;|z|^2)$ of $p=1=q$ confluent $k$-hypercats  for $k=5$ and $j=0,1,2,3,4$ (from thickest to thinest curves). We represent 
(a) Super-Poissonian $\alpha=1,\beta=4$, (b) Poissonian $\alpha=1=\beta$ and (c) Sub-Poissonian $\alpha=4,\beta=1$.}\label{mandelconflu}
\end{figure}

The Mandel parameter for the $k$-hypercat is explicitly
\be 
 \mathcal{Q}(\alpha,\beta;k,j;x)=x\left(\frac{ {}_p^k{F_q}^{j+2k-2}(\alpha,\beta;x) }{{ }_p^kF_q^{j+k-1}(\alpha,\beta;x)}-
 \frac{ {}_p^k{F_q}^{j+k-1}(\alpha,\beta;x) }{{ }_p^kF_q^j(\alpha,\beta;|x)}\right), \quad x=|z|^2.\label{mandelcat}
\ee
\begin{enumerate}
\item For  $f(nk+j) <f(nk+j-1)$ we have    $ \mathcal{Q}(\alpha,\beta;k,j;x)> 0$, 
which yields the  super-Poissonian distribution.
\item For  $f(nk+j-1) <f(nk+j)$  we have  $ \mathcal{Q}(\alpha,\beta;k,j;x)< 0 $
which yields the  sub-Poissonian distribution.
\end{enumerate}
In Figure \ref{mandelconflu} we represent the Mandel parameter \eqref{mandelcat} for confluent ($p=1=q$) $k$-hypercats for the super- sub- and Poissonian cases and 
$k=5$. We observe that $\mathcal{Q}(\alpha,\beta;k,0;0)=k-1$ and $\mathcal{Q}(\alpha,\beta;k,j;0)=-1$ for $j\not=0$. For large $x$, $\mathcal{Q}(\alpha,\beta;k,j;x)$ 
behaves like $\mathcal{Q}(\alpha,\beta,x)$, independently of $j$ for each $k$.

For completeness, in appendix \ref{meannf} we also  provide formulas for mean values, standard deviations and Mandel parameters 
for the nonlinear number operator $\hat{n}_f$. For higher-order squeezing of eigenstates of $\ah^k$ and eigenstates of  $\ah^k_f$ we address the reader 
to Refs. \cite{PRA46} and \cite{JPB35}, respectively.

The previous study indicates that the $k$-hypercat $|z;\alpha,\beta;k,j\rangle$ constitutes a good approximation to the number state $|j\rangle$ for high $k$ and small $|z|$. 
This fact suggests that there must be a circle representation for HCS, like the one in Eq.  \eqref{circlerepcan} for canonical CS, 
the expression \eqref{discretecircle}  being a discrete approximation. Let us study this fact in more detail.

\subsection{The circle representation for hypergeometric CS}\label{circleCS}

A circle representation \eqref{circlerepcan} of basis number states $|n\rangle$ in terms of an equally weighted superposition of canonical CS $|z\rangle$ on the circle of radius $r=|z|$ 
was proved in  \cite{PRA50}. Here we shall provide a variation of this formula but for  HCS \eqref{hypercs}. We shall state this result in the form of a Proposition.

\begin{proposition} A circle representation of basis number states $|n\rangle$ in terms of an equally weighted superposition of HCS $|z;\alpha,\beta\rangle$ on the circle of radius 
$r=|z|$ is given by
 \be
|n\rangle=\sqrt{{}_p{F}_q(\alpha,\beta; r^2)\,{}_p\rho_q(n)}\,\frac{r^{-n}}{2\pi}\int_0^{2\pi} e^{-in\theta}|r e^{i\theta};\alpha,\beta\rangle\, d\theta.\label{circlerephyper}
\ee
\end{proposition}
{\noindent \textbf{Proof:} Just multiply the expression \eqref{hypercs} by $z^{-m}$, integrate on the circle $C=\{z=re^{i\theta}\}$, for fixed radius $r$, and use the orthogonality 
property $\int_0^{2\pi} \exp(i\theta(n-m)) d\theta=2\pi \delta_{n,m}$. $\square$

In fact, this simple proof is again a consequence of the analytical nature of CS in general. 
For the present infinite-dimensional case, we can interpret \eqref{discretecircle} as a finite discretization of 
\eqref{circlerephyper}, in a similar sense to
\[\int_0^{2\pi}g(\theta)d\theta\simeq \frac{2\pi}{k}\sum_{l=0}^{k-1} g(2\pi l/k),\]
for a continuous function $g$ on the circle. The circle representation  also applies to the truncated finite-dimensional case.
Nevertheless, for the truncated case, a discrete finite exact  version of this circle representation exists. Let us discuss it in a separate subsection.

\subsection{\label{hyperlikecats} Exact finite circle representation for truncated $k$-hypercats}

As we have already noted in Section \ref{hypersec2}, HCS $|z;\alpha,\beta\rangle$ for some negative $\alpha$ or $\beta$ must be truncated 
[remember the form of the normalization factor in eq. \eqref{normfactN}] 
and, therefore, they are not anymore eigenstates of $\ah_f$. The same applies 
for $k$-hypercats. Nevertheless, we can still define truncated $k$-hypercats simply by mimicking expression \eqref{khypercats} and defining
\be
\label{reaaistrunc}
|z;\alpha,\beta;k,j\rangle={}_p^kF_q^j(\alpha,\beta;|z|^2)^{-\um}_N \sum_{n=0}^{(N-k+1)/k}\frac{z^{nk+j}}{\sqrt{{}_p\rho_q(nk+j)}} |nk+j\rangle,\quad j=0,1,\dots k-1,
\ee
where 
\be
{}_p^kF_q^j(\alpha,\beta;|z|^2)_N=\sum_{n=0}^{(N-k+1)/k}\frac{|z|^{2(nk+j)}}{|{{}_p\rho_q(nk+j)}|} 
\ee
is a normalization factor. Remember that $N$ is the absolute value of the 
largest negative $\alpha_j$ or $\beta_j$. Note that now $k$ must be a divisor of $N+1$.  

\begin{proposition}  The states \eqref{reaaistrunc} can 
also be written as an equally weighted superposition of truncated HCS as 
\bea
|z;\alpha,\beta;k,j\rangle =  \frac{1}{k}\left(\frac{{}_pF_q(\alpha,\beta;(-1)^\sigma|z|^2)_N}{{}_p^kF_q^j(\alpha,\beta;|z|^2)_N}\right)^{1/2}
\sum_{l=0}^{k-1}e^{-2\pi i jl/k}|ze^{2\pi i l/k };\alpha,\beta\rangle,\quad j=0, 1, \cdots k-1.\label{discretecircle2} 
\eea
Therefore, we shall call these states  ``truncated hypergeometric Schr\"odinger kittens'' or ``truncated $k$-hypercats'' for short. 
\end{proposition}
{\noindent \textbf{Proof:}    From the definition of truncated HCS in Section \ref{hypersec2}, we can define phase-shifted truncated HCS as
\begin{equation}
|ze^{2\pi i l/k};\alpha,\beta\rangle={}_pF_q(\alpha,\beta;(-1)^\sigma|z|^2)^{-\um}_N \sum_{n=0}^{N}\frac{z^n e^{2\pi i nl/k}}{\sqrt{{}_p\rho_q(n)}} |n\rangle, ,\quad l=0, 1, \cdots k-1.
\end{equation}
Multiplying both sides by $e^{-2\pi i jl/k}$, summing on $l$ and using the orthogonality condition \eqref{orthogonality}, we arrive to
\begin{equation}
\sum_{l=0}^{k-1}e^{-2\pi i jl/k}|ze^{2\pi i l/k };\alpha,\beta\rangle={}_pF_q(\alpha,\beta;(-1)^\sigma|z|^2)^{-\um}_N\,k  \sum_{m=0}^{(N-k+1)/k}\frac{z^{mk+j}}{\sqrt{{}_p\rho_q(mk+j)}} |mk+j\rangle.
\end{equation}
Comparing the right hand side with \eqref{reaaistrunc} we arrive to \eqref{discretecircle2}$\square$

For $k=N+1$, the sum in \eqref{reaaistrunc} has only one addend ($n=0$) and therefore $|z;\alpha,\beta;N+1,j\rangle=|j\rangle$; that is, 
the truncated $(N+1)$-hypercat in \eqref{reaaistrunc} coincides with the basis state 
$|j\rangle$. Therefore, for  $k=N+1$, the expression \eqref{discretecircle2} provides a discrete exact version of the circle representation \eqref{circlerephyper} for truncated HCS. 
Indeed, we make this explicit in a corollary

\begin{corollary}\label{discretecircprop}  A discrete exact circle representation of basis number states $|n\rangle$ in terms of an equally weighted superposition of 
truncated HCS  on the circle of radius $r=|z|$ is given by
\be
|n\rangle =   \frac{1}{N+1}\left(\frac{{}_pF_q(\alpha,\beta;(-1)^\sigma|z|^2)_N}{{}_p^{N+1}F_q^n(\alpha,\beta;|z|^2)_N}\right)^{1/2}
\sum_{l=0}^{N}e^{-2\pi i nl/(N+1)}|ze^{2\pi i l/(N+1) };\alpha,\beta\rangle,\quad n=0, 1, \cdots N.\label{discretecircle3} 
\ee
\end{corollary}

% \textbf{Statistics:} (we have to complete this)
% 
% Spin-$s$ $SU(2)$ CS ($p=0,\;q=1, \beta_1=-2s$): For $k=2,$ we have
% \bea 
% {}_0^2P_1^0(\cdot,-2s;m;|z|^2)&=&\frac{|z|^{2m}\;\;\delta_{m,0\, \mathrm{mod} \;2}}{m!\;   {}_{0}F_{3}\left(
% \begin{array}{c}
% \cdot\\
% -s+\frac{1}{2}, -s, \frac{1}{2}\end{array}\Big|\frac{|z|^4}{16}
% \right)_{2s}}, \nonumber\\  {}_0^2P_1^1(\cdot,-2s;m;|z|^2)&=&- \frac{2s\,|z|^{2(m-1)}\;\;\delta_{m,1 \,\mathrm{mod}\; 2}}{m!\;   {}_{0}F_{3}\left(
% \begin{array}{c}
% \cdot\\
% -s+1,-s+\frac{1}{2}, \frac{3}{2}\end{array}\Big|\frac{|z|^4}{16}
% \right)_{2s}}.
% \eea
% %\section{ Photon number statistics}

\section{Generating $k$-hypercats  by amplitude dispersion}\label{kerrsec}

As we have previously commented, standard eingenstates of $\hat{a}^k$ are non-classical states of light and can be generated via amplitude dispersion (see \cite{PRL57} for the initial proposal and 
\cite{Kirchmair} for recent experiments). To create and manipulate 
these multi-component Schr\"odinger cat states, useful in continuous variable quantum information protocols, a strong nonlinear interaction at the single photon level is required. A direct photon-photon interaction 
occurs in so-called Kerr media (a material whose refractive index depends on the intensity of the light). The anharmonic term in the Hamiltonian is taken to be proportional to $\hat{n}^\kappa,\kappa>1$. More precisely, 
consider  an anharmonic oscillator Hamiltonian of the form $H=\hbar\omega \hat{n}+\hbar\Omega\hat{n}^\kappa$, where $\hbar\omega$ is the energy-level splitting for the harmonic part and $\Omega$ is the strength of the 
anharmonic term (Kerr constant). In the interaction picture, an initial  CS $|z\rangle$  (describing for example a light beam traveling through such a material) acquires a phase shift $\phi_n(t)=\Omega t n^\kappa$, where 
$t$ is the interaction time of the light field with the material. Therefore, the revival time is $\tau=2\pi/\Omega$. 
Let us see that, at fractions $\tau/k$, the HCS evolves into a equally weighted superposition of HCS on a circle, with a similar structure to the 
$k$-hypercat. Indeed, the time evolution of a HCS \eqref{hypercs} in the interaction picture is 
\be |z,t;\alpha,\beta\rangle=e^ {-i\Omega t\hat{n}^\kappa}|z;\alpha,\beta\rangle={}_pF_q(\alpha,\beta;|z|^2)^{-\um}\sum_{n=0}^\infty 
\frac{z^n e^ {-i\phi_n(t)}}{\sqrt{{}_p\rho_q(n)}}|n\rangle.\label{hypercst}\ee 
At fractions $t_k=\tau/k=2\pi/(\Omega k)$ of the revival time we have $\exp(-i\phi_n(t_k))= \exp(-2\pi i /k)^{n^\kappa}$; these are powers of the $k$ roots of unity and the sequence is periodic. 
There are four different cases, according to the parity of $\kappa$ and $k$. Let us focus on the case $\kappa=2$, of which an experiment to engineer an artificial Kerr medium 
using a three-dimensional circuit quantum electro-dynamic architecture has been done in \cite{Kirchmair}. For this case, writting $n=n'k+j$, the evolved state \eqref{hypercst} 
at time $t_k$  can be written as a superposition of $k$-hypercats \eqref{khypercats} as
\be |z,t_k;\alpha,\beta\rangle={}_pF_q(\alpha,\beta;|z|^2)^{-\um}\sum_{j=0}^{k-1} \sum_{n'=0}^\infty 
\frac{z^{n'k+j} e^ {-2\pi i j^2/k}}{\sqrt{{}_p\rho_q(n'k+j)}}|n'k+j\rangle= \sum_{j=0}^{k-1}\frac{{}_p^kF_q^j(\alpha,\beta;|z|^2)^{\um}}{{}_pF_q(\alpha,\beta;|z|^2)^{\um}}e^ {-2\pi i j^2/k}|z;\alpha,\beta;k,j\rangle.\label{hypercstk}\ee
Now, using the expression \eqref{discretecircle}, we can write the previous state as an equally weighted superposition of HCS on the circle of radius $|z|$ as follows
\be
 |z,\tau/k;\alpha,\beta\rangle=\frac{1}{k}\sum_{j=0}^{k-1}\sum_{l=0}^{k-1}e^{-2\pi ij(j+l)/k}|ze^{2\pi i l/k};\alpha,\beta\rangle .\label{hypercst2}\ee 
In order to visualize the structure of this state in phase space, we shall make use of the lower or covariant symbol 
$Q_\psi^{\alpha,\beta}(z)=|\langle \psi|z; \alpha,\beta\rangle|^2$ of the density matrix $\rho=|\psi\rangle\langle\psi|$. This lower symbol coincides with the well known Husimi $Q$-function for the particular case 
of $|z; \alpha,\beta\rangle=|z;\cdot,\cdot\rangle$, that is, canonical CS. Then we shall refer to $Q_\psi^{\alpha,\beta}(z)$ as ``generalized Husimi $Q$-function''. 

In our case, using the HCS overlap \eqref{hypercsoverlap}, the generalized Husimi $Q$-function of the state $|\psi\rangle=|z_0,\tau/k;\alpha,\beta\rangle$ in \eqref{hypercst2} is 
\be
Q_{z_0,k}^{\alpha,\beta}(z)=|\langle z_0,t_k;\alpha,\beta|z;\alpha,\beta\rangle|^2=\left| \frac{1}{k}{\sum _{j=0}^{k-1} \sum _{l=0}^{k-1} \frac{\exp \left(\frac{2 \pi  i j (j+l)}{k}\right) 
\, _1F_1\left(\alpha ;\beta ;\bar{z}_0 \exp \left(-\frac{2 \pi  i l}{k}\right) z\right)}{\sqrt{\, _1F_1\left(\alpha ;\beta ;|z_0|^2\right) 
\, _1F_1\left(\alpha ;\beta ;|z|^2\right)}}}\right|^2.\label{husifunc}
\ee
Let us have a closer look to the structure of the evolved state $|z_0,\tau/k;\alpha,\beta\rangle$. Note that it can also be written as
\be
 |z_0,\tau/k;\alpha,\beta\rangle=\frac{1}{k} \sum_{l=0}^{k-1} e^{i\pi \frac{l^2}{2k}} \left(  \sum_{j=0}^{k-1}e^{-2\pi i(j+\frac{l}{2})^2/k}\right)   |z_0e^{2\pi i l/k};\alpha,\beta\rangle.\label{hypercst3}\ee 
The sum in $j$ can be performed explicitly, and three different cases appear:
 \begin{itemize}
  \item[a)] $k$ is odd: All $k$ values of $l$ contribute, therefore there are $m=k$ components in the sum, distributed as the $m$-th roots of unity multiplied by $z_0$.
  \item[b)] $k$ is multiple of 4: Only even values of $l$ contribute, therefore there are $m=k/2$ components in the sum, distributed as the $m$-th roots of unity multiplied by $z_0$.
  \item[c)] $k$ is even but not multiple of 4:  Only odd values of $l$ contribute, therefore there are $m=k/2$ components in the sum, distributed as the $m$-th roots of unity multiplied by $z_0$ and rotated by $e^{i \frac{2\pi}{k}}$.
 \end{itemize}
In \cite{Kirchmair} it is commented that to distinguish the $m$ components of a (canonical) kitten, the CSs  have to be separated by more
than twice their width on a circle with a radius given by the initial displacement $r=|z_0|$. In other words, the coherent states have to be quasi-orthogonal, which means that $2\pi r/m$ has to be large enough. 
For a canonical CS, the standard deviation of the Gaussian is $\sigma=1$ (standard normal distribution). We know that 
about 99.7\% of values drawn from a normal distribution are within three standard deviations (``3-$\sigma$ rule''), which gives $2\pi r/m\gtrapprox 3\sigma$.  
This means that, for a displacement of $r=|z_0|=2$,  the maximum number of canonical CS that can be distinguished is about $m\approx 4$ (thus $k=8$). 
Let us compare the temporal evolution of canonical CS with that of HCS of the same family. 
For example, we know that the confluent hypergeometric function ${}_1F_1(\alpha,\alpha;|z|^2)=\exp(|z|^2)$ and, therefore, canonical CS is a particular case of $p=1=q$ HCS 
$|z;\alpha,\beta\rangle={}_1F_1(\alpha,\beta;|z|^2)^{-\um}{}_1F_1(\alpha,\beta;z \ad)|0\rangle$. Let us consider the same three cases as we did at the end of Section \ref{statprop}, that is: 
1) Poissonian $\alpha=\beta$, 2) Super-Poissonian $\alpha<\beta$ and  3) Sub-Poissonian $\alpha> \beta$. The corresponding 
(normalized) probability distributions are $f_{\alpha,\beta}(x)={}_1F_1(\alpha,\beta;x^2)^{-\um}/\mathcal{N}_{\alpha,\beta}$, where $\mathcal{N}_{\alpha,\beta}=\int_{-\infty}^\infty {}_1F_1(\alpha,\beta;x^2)^{-\um}dx$ is a normalization factor. 
The variance (squared standard deviation) is  $\sigma_{\alpha,\beta}^2=\int_{-\infty}^\infty x^2 f_{\alpha,\beta}(x)dx$. For example $\sigma_{1,1}=1, \sigma_{1,3}\simeq 1.32$ and $\sigma_{3,1}\simeq 0.75$. 
Note that $\sigma_{1,3}>\sigma_{1,1}>\sigma_{3,1}$. Using the rough estimate $m\lessapprox 2\pi r/(3\sigma)$ we realize that, for a displacement of $r=2$,  
the maximum number of HCS that can be distinguished from the multicomponent hypercat \eqref{hypercst2}  is about $m\approx 5$ (thus $k=5$ or $k=10$) for $\sigma_{3,1}$ (sub-Poissonian case) 
and $m\approx 3$ (thus $k=3$ or $k=6$) for $\sigma_{1,3}$ (super-Poissonian).  Indeed, in Figure \ref{Husicat} we represent the generalized Husimi $Q$-function $Q_{z_0,k}^{\alpha,\beta}(z)$ of $|z_0,\tau/k;\alpha,\beta\rangle$ for $k=1,15,8,6,5,4,3,2$ and $z_0=2$. 
Let us analyze each case separately: 
\begin{itemize}

 \item For $k=1$ (first column), the state 
$|z_0,\tau;\alpha,\beta\rangle$ coincides with the initial state $|z_0;\alpha,\beta\rangle$ (complete revival). 
 \item For $k=15$ (second column), the number of components $m=15$ is greater than the three critical values 
$m\simeq 3, 4$ and $5$ for super- sub- and Poissonian cases, respectively, and therefore the 15 components can not be distinguished. 
 \item For $k=8$ (third column), the $m=4$ components can be clearly distinguished for the Poissonian and sub-Poissonian states (second and third rows, respectively), but not really for the super-Poissonian state (first row), 
which requires $m\lessapprox 3$. 
 \item For $k=6$ (fourth column), the $m=3$ components can be  distinguished in the three cases, the sub-Poissonian case improving the Poissonian and super-Poissonian cases. Note the similarities between 
the cases $k=6$ (fourth column) and $k=3$ (seventh column) which, according items a) and c) after \eqref{hypercst3}, the 3 components are distributed as the $3$-th roots of unity multiplied by $z_0$, the case $k=6$ being rotated an angle 
$\pi/3$ with respect to the case $k=3$. 
 \item For $k=5$ (fifth column), the $m=5$ components can be distinguished for the sub-Poissonian state, but not really for the Poissonian and super-Poissonian states. 
 \item For $k=4$ (sixth column), the $m=2$ components can be  distinguished in the three cases.  
 \item Finally, for $k=2$ the initial state $|z_0;\alpha,\beta\rangle$ goes to the parity reversed $|-z_0;\alpha,\beta\rangle$. 
\end{itemize}
This analysis corroborates the multicomponent structure of the state \eqref{hypercst3} and the rough estimate $m\lessapprox 2\pi r/(3\sigma)$ for the number of distinguishable components as a function of the standard 
deviation $\sigma$ and the initial displacement $r$.

\begin{figure}[h]
\begin{center}
(a)\includegraphics[width=15cm]{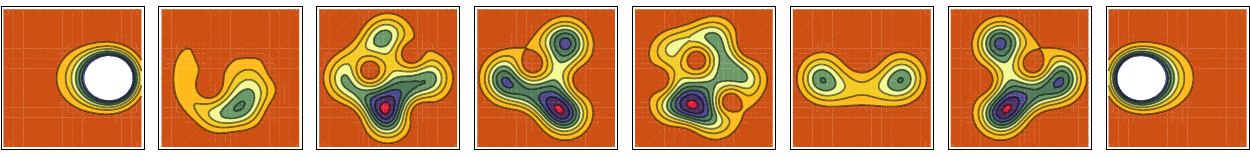}\\ 
(b)\includegraphics[width=15cm]{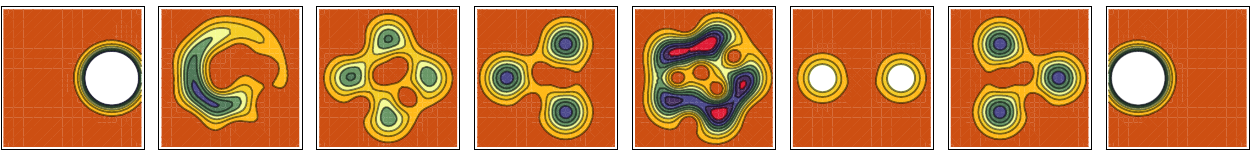}\\ 
(c)\includegraphics[width=15cm]{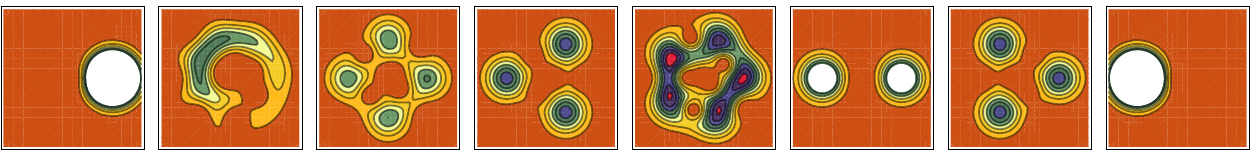}
\end{center}
\caption{Contour plots of the generalized Husimi $Q$-function $Q_{z_0,k}^{\alpha,\beta}(z)$ \eqref{husifunc} of the multicomponent 
state \eqref{hypercst2} generated by temporal evolution in a Kerr medium from a confluent ($p=1=q$) HCS with $z_0=2$ at times $t=0$ and 
$t_k/\tau=1/k=1/15,1/8,1/6,1/5,1/4,1/3,1/2$ (from left to right). The first row (a) corresponds to the super-Poissonian $\alpha=1,\beta=3$ case, the second row (b) corresponds to the Poissonian  
$\alpha=1,\beta=1$ case and the third row (c) to the sub-Poissonian case $\alpha=3,\beta=1$. We show the window $|x|,|y|\leq 3.5$ for $z=x+i y$. }\label{Husicat}
\end{figure}

\section{Conclusions and outlook}\label{conclusec}

We have studied mathematical and statistical properties of circular multiphoton CS  of hypergeometric type (``$k$-hypercats''). They constitute a discrete version of the circle representation and they interpolate 
between number and coherent states, with a critical displacement $z_0$ separating the number from de CS region/behavior. These  macroscopic equally-weighted superpositions of quantum states 
(also known as  multicomponent Schr\"odinger cat states, generalizing the usual  two-component, even/odd Schr\"odinger cats)  exhibit interesting non-classical properties and 
can be generated  by  amplitude dispersion in a Kerr medium. We analyzed the structure in phase space of their generalized Husimi $Q$-function in the super- and sub-Poissonian cases at different fractions of the revival time. 
This analysis corroborates their multicomponent structure.

Actual quantum technologies allow to physically realize different kinds of nonlinear $f$-CS and their quantum superpositions. Since the initial laser-driven trapped ion proposal \cite{MatosVogel}, many more generation schemes 
have been explored. The literature is huge and we address the reader to, for example, Ref. \cite{YanZhuLi} for the generation of $f$-CS of the mechanical resonator in an optomechanical microcavity, and references therein on other 
generation schemes. Of special interest are actual generation schemes inspired in two-dimensional (Dirac-like) materials like graphene, in particular, the use of nanomechanical graphene resonators to prepare vibrational quantum states 
\cite{graphene2} and to generate  macroscopic superposition states \cite{graphene3}. The mathematical construction of Barut-Girardello graphene CS has been recently studied in \cite{graphene1}. The extension to other 
two-dimensional allotropes (like silicene, exhibiting topological insulator properties) could introduce some novelties. This is work in progress.

%%%%%%%%%Phase op
% 
% \bibitem{Susskind} L. Susskind and J. Glogower, Quantum mechanical phase and time operator, Physics \textbf{1} (1964)
% 49
% \bibitem{Nieto} P. Carruthers and M.M. Nieto, %Phase and Angle Variables in Quantum Mechanics, 
% Rev. Mod. Phys. \textbf{40}
% (1968) 411-440
% \bibitem{Pegg1} D.T. Pegg and S.M. Barnett, %Unitary Phase Operator in Quantum Mechanics, 
% Europhys. Lett. \textbf{6}
% (1988) 483-487
% \bibitem{Pegg2} D.T. Pegg and S.M. Barnett, %Phase Properties of the Quantized Single-mode Electromagnetic Field,  
% Phys. Rev. \textbf{A39} (1989)
% 1665-1675
% \bibitem{JPA31} J.A. Gonz\'alez and M. A del Olmo, %Coherent states on the circle, 
% J. Phys. A: Math. Gen. \textbf{31} (1998) 8841-8857.
% \bibitem{GazeauPLA} P.L. García de León and J.P. Gazeau, %Coherent state quantization and phase operator, 
% Physics Letters \textbf{A361} (2007) 301-304
% 

\appendix

\section{Coherent states on the circle}\label{CScircle}
Let us make a clarification. Do not confuse the circle representation with ``CS on the circle''. 
This case could be seen as an extension of Susskind-Glogower operators \eqref{SG} by letting $n$ 
to run on $\mathbb{Z}$. In this case, we recover the phase operators for the Euclidean group $E(2)$:
\begin{equation}
\hat{a}_f=\hat{U}=\sum_{n=-\infty}^\infty |n\rangle\langle n+1|\,,\qquad
\hat{a}^\dag_f=\hat{U}^\dag=\sum_{n=-\infty}^\infty |n+1\rangle\langle n|  \,,
\end{equation}
satisfying $\hat{U}|n\rangle = |n-1\rangle$ and $\hat{U}^\dagger|n\rangle = |n+1\rangle\,,n\in \mathbb{Z}$. These $E(2)$ phase operators satisfy 
$\hat{U}\hat{U}^\dagger=  \hat{U}^\dagger\hat{U}  = \hat I$, and therefore they are unitary quantum phase operators. 
They were introduced by Louisell \cite{Louisell}, and further studied in \cite{Newton}. From the group-theoretical point of view, $\hat{U}$ and $\hat{U}^\dag$,
which are commuting operators, close a finite-dimensional Lie-algebra with the number operator $\hat{n}$, the Euclidean algebra $E(2)$. 
In this context, $\hat{n}$ can be understood as rotations on the circumference $-i\frac{\partial\, }{\partial \phi}$ and $\hat{U},\hat{U}^\dag$ 
as multiplication by a phase $e^{-i\phi},\,e^{i\phi}$. Nonlinear $f$-CS for this case  leads to
\begin{equation}
 |z,f\rangle \equiv |e^{-in\phi}\rangle = (2\pi)^{-1/2}\sum_{n=-\infty}^\infty e^{-in\phi} |n>
\end{equation}
since in this case $z$ is restricted to the circle $|z|=1$. See \cite{Newton} for the properties of these states. Note that Pegg \& Barnett phase operators are constructed from states obtained 
by considering a finite sum  from $n=0$ up to $N$ in the previous equation and restricting $\phi$ to the $N$th-roots of unity.

Perelomov type coherent states can also be introduced for the case of $E(2)$, but since its representations are not square-integrable \cite{deBievre,IshamKaluder}, different approaches have been used to
define suitable coherent states. In \cite{deBievre} (see also \cite{Torresani} for applications on signal processing on the circle) the parameters of the $E(2)$ group are restricted to a cylinder, i.e. 
the phase space of the circle, and even in this case suitable (admissibility) conditions
must be imposed on the fiducial vector. See \cite{GazeauPLA} and \cite{GazeauCircle2018} for a recent account of these states.

There are other approaches for the definition of coherent states on the circle, which are not related with the Susskind-Glogower operators nor with the $E(2)$ groups, 
like \cite{deBievreGonzalez,JPA31}. See the review \cite{Kastrup} for details on these and other families of coherent states on the circle.

\section{Normalization factor of $k$-hypercats\label{apA}}
 For $j=1, 2, \cdots, k-1$, the  normalization factor   
\[{}_p^kF^j_q(\alpha,\beta;x)= \sum_{n=0}^\infty\frac{x^{nk+j}}{{}_p\rho_q(nk+j)}, j=0,1, 2, \cdots, k-1,\;x=|z|^2 \]  
has the following expression
\be
\label{expj}
 {}_p^kF^j_q(\alpha,\beta;x)=\frac{(\alpha_1, \alpha_2, \cdots, \alpha_p)_j}{(\beta_1, \beta_2, \cdots, \beta_q)_j}\frac{x^j}{j!}
  {}_{pk+1}F_{(q+1)k}\left(
\begin{array}{c}
\Delta_k\left[\frac{\alpha_1+j}{k}\right],    \cdots, \Delta_k\left[\frac{\alpha_p+j}{k}\right], \Delta_1\Big[1\Big]\\
\Delta_k\left[\frac{\beta_1+j}{k}\right], \cdots, \Delta_k\left[\frac{\beta_q+j}{k}\right], \Delta_{k}\left[\frac{j+1}{k}\right]
\end{array}\Bigg|(xk^{p-q-1})^k
\right),
\ee
where $\displaystyle \Delta_i\left[\frac{a}{b}\right]=  \frac{a}{b}, \frac{a+1}{b}, \cdots,  
\frac{a+i-1}{b}$ and $\displaystyle \Delta_1\Big[1\Big]=1.$ Indeed
\be
\label{resasd}
{}_p^kF^j_q(\alpha,\beta;x)= \sum_{n=0}^\infty\frac{x^{nk+j}}{ {}_p\rho_q(nk+j)}= \sum_{n=0}^\infty\frac{(\alpha_1, \alpha_2, \cdots, \alpha_p)_{nk+j}}{(\beta_1, \beta_2, \cdots, \beta_q)_{nk+j}}\frac{ \,x^{nk}}{(nk+j)!}.
\ee
Rewriting 
\bea
&&(nk+j)!=j!(j+1)_{nk}=j!k^{nk}\left(\frac{j+1}{k}, \frac{j+2}{k}, \cdots, \frac{j+k}{k}\right)_n,\\
 &&(\alpha_1)_{nk+j}=(\alpha_1)_j(\alpha_1+j)_{nk}=k^{nk}(\alpha_1)_j\left(\frac{\alpha_1+j}{k}, \frac{\alpha_1+j+1}{k}, \cdots, \frac{\alpha_1+j+k-1}{k}\right)_n
\eea
the expression (\ref{resasd}) takes the form
\be
\label{rsd}
 {}_p^kF^j_q(\alpha,\beta;x)=\frac{(\alpha_1, \alpha_2, \cdots, \alpha_p)_{j}}{(\beta_1, \beta_2, \cdots, \beta_q)_{j}}\frac{ \,x^{j}}{j!} \sum_{n=0}^\infty\frac{\left(\frac{\alpha_1+j}{k}, \cdots,  \frac{\alpha_1+j+k-1}{k},\cdots, \frac{\alpha_p+j}{k}, \cdots,  \frac{\alpha_p+j+k-1}{k}\right)_n\,\Big[(xk^{p-q-1})^k\Big]^{n}}{\left(\frac{\beta_1+j}{k}, \cdots,  \frac{\beta_1+j+k-1}{k},\cdots, \frac{\beta_q+j}{k}, \cdots,  \frac{\beta_q+j+k-1}{k},  \frac{j+1}{k}, \cdots,  \frac{j+k}{k}\right)_n}
\ee
If we multiply and divide by $n!$ inside the sum, then this expression can be written like  \eqref{expj}.
Note that for $j=0,$ the  normalization factor reduces to
\be
\label{expj0}
 {}_p^kF^0_q(\alpha,\beta;x)=
  {}_{pk}F_{(q+1)k-1}\left(
\begin{array}{c}
\Delta_k\left[\frac{\alpha_1}{k}\right],    \cdots, \Delta_k\left[\frac{\alpha_p}{k}\right]\\
\Delta_k\left[\frac{\beta_1}{k}\right], \cdots, \Delta_k\left[\frac{\beta_q}{k}\right], \Delta_{k-1}\left[\frac{1}{k}\right]
\end{array}\Bigg|(xk^{p-q-1})^k
\right).
\ee

\section{Mean values, standard deviations and Mandel parameters of $\hat{n}_f$ in a $k$-hypercat}\label{meannf}
For  $k\geq 3$,  we have 
\be
\langle \hat{n}_f\rangle(\alpha,\beta;j;|z|^2)=|z|^2\;\frac{{}_p^k{F_q}^{j-1}(\alpha,\beta;|z|^2) }{{ }_p^kF_q^j(\alpha,\beta;|z|^2)}
\ee
and 
\be
\sigma_{\hat{n}_f}=\sqrt{|z|^4\;\frac{{}_p^k{F_q}^{j-2}(\alpha,\beta;|z|^2) }{{ }_p^kF_q^j(\alpha,\beta;|z|^2)}+|z|^2\;\frac{{}_p^k{F_q}^{j-1}(\alpha,\beta;|z|^2) }{{ }_p^kF_q^j(\alpha,\beta;|z|^2)}-
\left[ |z|^2\;\frac{{}_p^k{F_q}^{j-1}(\alpha,\beta;|z|^2) }{{ }_p^kF_q^j(\alpha,\beta;|z|^2)}\right]^2},
\ee
which gives the Mandel parameter
\be 
\label{:mandf}
 \mathcal{Q}_f(\alpha,\beta;k,j;x)=x\left(\frac{ {}_p^k{F_q}^{j-2}(\alpha,\beta;x) }{{ }_p^kF_q^{j-1}(\alpha,\beta;x)}-
 \frac{ {}_p^k{F_q}^{j-1}(\alpha,\beta;x) }{{ }_p^kF_q^j(\alpha,\beta;|x)}\right), \quad x=|z|^2.
\ee

\newpage

\thispagestyle{empty}

\begin{sidewaystable}[hb]

\centering
  \resizebox{0.85\textheight}{!}{\begin{minipage}{\textwidth}
     %   \caption{Table caption}
\hspace*{-5cm}
\begin{tabular}{|c|c|c|c|c|c|c|c|c|c|c|}
\hline
$(p,q)$ & $(\alpha,\beta)$  & $f(n)$ & $\ah_f$ & $|z;\alpha,\beta\rangle$  & $\mathcal{N}_f(|z|)$ & Domain & Related group & Type & Known as  &Dual  \\
\hline
$(0,0)$ & $(\cdot,\cdot)$ & 1&  $\ah_f|n\rangle=\sqrt{n}|n-1\rangle$ & $|z;\cdot,\cdot\rangle=e^{-|z|^2/2}\sum_{n=0}^\infty \frac{z^n}{\sqrt{n!}}|n\rangle$ & $e^{|z|^2}$ & $z\in \mathbb{C}$ & H-W & GP,BG & Canonical CS & Self-dual \\
\hline
$(1,0)$ & $(2s,\cdot)$ &  $\frac{1}{\sqrt{2s+n-1}}$   & $\ah_f|n\rangle=\sqrt{\frac{n}{2s+n-1}}|n-1\rangle$ & $|z;2s,\cdot\rangle=(1-|z|^2)^s\sum_{n=0}^\infty \sqrt{\binom{2s+n-1}{n}}{z^n}|n\rangle$ & 
$(1-|z|^2)^{-2s}$ & $|z|<1$  &SU(1,1) & GP &  GP SU(1,1) CS & BG SU(1,1) CS\\
\hline
$(0,1)$ & $(\cdot,2s)$ & $\sqrt{2s+n-1}$ & $\ah_f|n\rangle=\sqrt{n(2s+n-1)}|n-1\rangle$ & $|z;\cdot,2s\rangle={}_0F_1(\cdot,2s;|z|^2)^{-1/2}\sum_{n=0}^\infty \binom{2s+n-1}{n}^{-1/2}\frac{z^n}{n!}|n\rangle$ & 
${}_0F_1(\cdot,2s;|z|^2)$ & $z\in \mathbb{C}$ &SU(1,1) & BG &  BG SU(1,1) CS & GP SU(1,1) CS \\
\hline
$(1,0)$ & $(1,\cdot)$ &  $\frac{1}{\sqrt{n}}$   & $\ah_f|n\rangle=|n-1\rangle$ & $|z;1,\cdot\rangle = \sqrt{1-|z|^2} \sum_{n=0}^\infty z^n |n\rangle$ & 
$\frac{1}{1-|z|^2}$ & $|z|<1$  &  & BG & Susskind-Glogower CS & Dual SG CS \\
\hline
$(0,1)$ & $(\cdot,1)$ &  $\sqrt{n}$   & $\ah_f|n\rangle=n |n-1\rangle$ & $|z;\cdot,1\rangle = I_0(2|z|)^{-1/2} \sum_{n=0}^\infty \frac{z^n}{n!} |n\rangle$ & 
$I_0(2|z|)$ & $z\in \mathbb{C}$  &  & BG & Dual SG CS & SG CS \\
\hline
$(2,0)$ & $((1,1),\cdot)$ &  $\frac{1}{n}$   & $\ah_f|n\rangle=\frac{1}{\sqrt{n}}|n-1\rangle$ & $|z;(1,1),\cdot\rangle={}_2F_0((1,1),\cdot;|z|^2)^{-1/2}\sum_{n=0}^\infty \sqrt{n!}z^n  |n\rangle$ & 
   ${}_2F_0((1,1),\cdot;|z|^2)=\infty$ & $|z|=0$  &H-W &  &  Inverse bosonic CS  &Dual Inverse bosonic CS \\
   \hline
$(0,2)$ & $(\cdot,(1,1))$ &  $n$   & $\ah_f|n\rangle=n^{3/2}|n-1\rangle$ & $|z;\cdot,(1,1)\rangle={}_0F_2(\cdot,(1,1);|z|^2)^{-1/2}\sum_{n=0}^\infty \frac{z^n}{(n!)^{3/2}}  {z^n}|n\rangle$ & 
   ${}_0F_2(\cdot,(1,1);|z|^2)$ & $z\in \mathbb{C}$  &H-W &  & Dual Inverse bosonic CS  &Inverse bosonic CS \\
\hline
\end{tabular}
 \caption{Hypergeometric CS}% Add 'table' caption
 \label{TableHyperCS}
 
 \vspace{2cm}
 
\hspace*{-5cm}
 \begin{tabular}{|c|c|c|c|c|c|c|c|c|c|c|}
\hline
$(p,q)$ & $(\alpha,\beta)$  & $f(n)$ & $\ah_f$ & $|z;\alpha,\beta\rangle$  & $\mathcal{N}_f(|z|)$ & Domain & Related group & Type & Known as  &Dual  \\
 \hline
$(0,1)$ & $(\cdot,-2s)$ & $\sqrt{(n-2s-1}$ & $\ah_f=i\sqrt{n(n-2s-1)}|n-1\rangle$ & $|z;\cdot,-2s\rangle={}_0F_1(\cdot,-2s;-|z|^2)_{2s}^{-1/2}\sum_{n=0}^{2s} \binom{2s}{n}^{-1/2}\frac{(iz)^n}{n!}|n\rangle$ & 
${}_0F_1(\cdot,-2s;-|z|^2)_{2s}$ & $z\in \mathbb{C}$ &SU(2) & BG&  BG SU(2) CS & GP SU(2) CS \\
\hline
$(1,0)$ & $(-2s,\cdot)$ &  $\frac{1}{\sqrt{n-2s-1}}$   & $\ah_f|n\rangle=-i\sqrt{\frac{n}{n-2s-1}}|n-1\rangle$ & $|z;-2s,\cdot\rangle=(1+|z|^2)^{-s}\sum_{n=0}^{2s}\binom{2s}{n}^{1/2}(iz)^n|n\rangle$ & 
$(1+|z|^2)^{2s}$ & $z\in \mathbb{C}$  &SU(2) & GP &  GP SU(2) CS & BG SU(1,1) CS\\
\hline
% $(1,0)$ & $\{(1),\cdot\}$ &  $\frac{1}{\sqrt{n}}$   & $\ah_f|n\rangle=|n-1\rangle$ & $|z;1,\cdot\rangle = \sqrt{1-|z|^2} \sum_{n=0}^\infty z^n |n\rangle$ & 
% $\frac{1}{1-|z|^2}$ & $|z|<1$  &  & BG & Susskind-Glogower CS & Dual SG CS \\
% \hline
% $(0,1)$ & $\{\cdot,(1)\}$ &  $\sqrt{n}$   & $\ah_f|n\rangle=n |n-1\rangle$ & $|z;\cdot,1\rangle = I_0(2|z|)^{-1/2} \sum_{n=0}^\infty \frac{z^n}{n!} |n\rangle$ & 
% $I_0(2|z|)$ & $z\in \mathbb{C}$  &  & BG & Dual SG CS & SG CS \\
% \hline
\end{tabular}
 \caption{Truncated Hypergeometric CS}% Add 'table' caption
\label{TableTruncHyperCS}

\end{minipage}} 
 
\end{sidewaystable}

\section*{Acknowledgements}

SA thanks   MC and JG for their hospitality during his stay at the University of Granada where this
work was done, and the  Coimbra Group for the financial
support. This study  has been partially financed by the Consejer\'\i a de Conocimiento, Investigaci\'on y Universidad, 
Junta de Andaluc\'\i a and European Regional Development Fund (ERDF) under projects with Refs. FQM381 and SOMM17/6105/UGR, and by the Spanish MICINN under project PGC2018-097831-B-I00. 
JG thanks the Spanish MICINN for financial support (FIS2017-84440-C2-2-P).

%%%%%%%%%%%%%%%%%% References
% %%%%%%%%%%%%%%%%%%%%%%%%%%%%%%%%%%%%%%%%%%%%%%%%%%%%%%%%%%%%%%%%%%%%%%%
% \def\JMP #1 #2 #3 {J. Math. Phys. {\bf#1},\ #2 (#3)}
% \def\JPA #1 #2 #3 {J. Phys. A {\bf#1},\ #2 (#3)}
% \def\JPD #1 #2 #3 {J. Phys. D {\bf#1},\ #2 (#3)}
% \def\PRL #1 #2 #3 { Phys. Rev. Lett. {\bf#1},\ #2 (#3)}
% \def\PR #1 #2 #3 { Phys. Rev. {\bf#1},\ #2 (#3)}
% \def\PLA #1 #2 #3 { Phys. Lett. A {\bf#1},\ #2 (#3)}
% \def\PLB #1 #2 #3 { Phys. Lett. B {\bf#1},\ #2 (#3).}
% \def\PRD #1 #2 #3 { Phys. Rev. D {\bf#1},\ #2 (#3).}
% \def\PRA #1 #2 #3 { Phys. Rev. A {\bf#1},\ #2 (#3).}
% 
% %%%%%%%%%%%%%%%%%%%%%%%%%%%%%%%%%%%%%%%%%%%%%%%%%%%%%%%%%%%%%%%%%%%%%%%

%%%%%%%%%%%%%%%%%%%%%%%%%%%%%%%%%%%%%%%%%%%%%%%%%%%%%%%%%%%%%%%%%%%%%%%
% volume page year
\def\JMP #1 #2 #3 {J. Math. Phys. {\bf#1}\  (#3) \ #2 }
\def\JPA #1 #2 #3 {J. Phys. A {\bf#1}\ (#3) \ #2 }
\def\JPB #1 #2 #3 {J. Phys. B {\bf#1}\ (#3) \ #2 }
\def\JPD #1 #2 #3 {J. Phys. D {\bf#1}\ (#3) \ #2 }
\def\PRL #1 #2 #3 {Phys. Rev. Lett. {\bf#1}\ (#3)\ #2 }
\def\PR #1 #2 #3 { Phys. Rev. {\bf#1}\ (#3)\ #2 }
\def\PL #1 #2 #3 { Phys. Lett.  {\bf#1} (#3) \ #2 }
\def\PLA #1 #2 #3 { Phys. Lett. A {\bf#1} (#3) \ #2 }
\def\PLB #1 #2 #3 { Phys. Lett. B {\bf#1}\ (#3)\ #2 }
\def\PRA #1 #2 #3 { Phys. Rev. A {\bf#1}\ (#3)\ #2 }
\def\PRB #1 #2 #3 { Phys. Rev. B {\bf#1}\ (#3)\ #2 }
\def\PRD #1 #2 #3 { Phys. Rev. D {\bf#1}\ (#3)\ #2 }
\def\AP #1 #2 #3 {Ann. Phys. (NY) {\bf#1}\ (#3) \ #2 }
\def\CMP #1 #2 #3 {Comm. Math. Phys.  {\bf#1}\ (#3) \ #2 }
\def\JMP #1 #2 #3 {J. Math. Phys.  {\bf#1}\ (#3) \ #2 }
\def\PS #1 #2 #3 {Physica Scripta  {\bf#1}\ (#3) \ #2 }
\def\Nat #1 #2 #3 {Nature  {\bf#1}\ (#3) \ #2 }
\def\NatP #1 #2 #3 {Nature Physics {\bf#1}\ (#3) \ #2 }
\def\JFAA #1 #2 #3 {J Four. Anal. Appl.  {\bf#1}\ (#3) \ #2 }
\def\EL #1 #2 #3 {Europhys. Lett.  {\bf#1}\ (#3) \ #2 }
\def\EPL #1 #2 #3 {EPL  {\bf#1}\ (#3) \ #2 }

\def\CPL #1 #2 #3 {Chin. Phys. Lett.  {\bf#1}\ (#3) \ #2 }
\def\JOB  #1 #2 #3 {J. Opt. B  {\bf#1}\ (#3) \ #2 }
\def\OL #1 #2 #3 {Opt. Lett.  {\bf#1}\ (#3) \ #2 }
\def\SP #1 #2 #3 {Signal Processing  {\bf#1}\ (#3) \ #2 } 
\def\NP #1 #2 #3 {Nucl. Phys.  {\bf#1}\ (#3) \ #2 } 

\def\Phys #1 #2 #3 {Physica  {\bf#1}\ (#3) \ #2 }
\def\JRLR  #1 #2 #3 {J. Russ. Laser Res.  {\bf#1}\ (#3) \ #2 }
\def\AMP #1 #2 #3 {Advances in Mathematical Physics  {\bf#1}\ (#3) \ #2 }
\def\CEJP #1 #2 #3 {Central European Journal of Physics  {\bf#1}\ (#3) \ #2 } 
\def\EPJP #1 #2 #3 {Eur. Phys. J. Plus  {\bf#1}\ (#3) \ #2 } 

%%%%%%%%%%%%%%%%%%%%%%%%%%%%%%%%%%%%%%%%%%%%%%%%%%%%%%%%%%%%%%%%%%%%%%%

\end{document}